\documentclass[preprint,authoryear,12pt]{elsarticle}

\usepackage{hyperref}
\usepackage{amssymb, amsmath, multirow, color}

\journal{Astronomy \& Computing}

\newcommand{\ddt}{\frac{\partial}{\partial t}}
\newcommand{\ddr}{\frac{\partial}{\partial r}}

\newcommand{\bmath}[1]{\ensuremath{\boldsymbol{#1}}}
\DeclareMathAlphabet{\textbfss}{\encodingdefault}{\sfdefault}{bx}{n}
\newcommand{\red}[1]{#1}

\begin{document}

\begin{frontmatter}

\title{\texttt{VADER}: A Flexible, Robust, Open-Source Code for Simulating Viscous Thin Accretion Disks}

\author{Mark R.~Krumholz and John C.~Forbes}
\address{Astronomy Department, University of California, Santa Cruz, 95064, USA; mkrumhol@ucsc.edu and jcforbes@ucsc.edu}


\begin{abstract}
The evolution of thin axisymmetric viscous accretion disks is a classic problem in astrophysics. While \red{models based on this simplified geometry} provide only approximations to the true processes of instability-driven mass and angular momentum transport, their simplicity makes them invaluable tools for both semi-analytic modeling and simulations of long-term evolution where two- or three-dimensional calculations are too computationally costly. Despite the utility of these models, \red{the only} publicly-available framework\red{s} for simulating them \red{are rather specialized and non-general}. Here we describe a highly flexible, general numerical method for simulating viscous thin disks with arbitrary rotation curves, viscosities, boundary conditions, grid spacings, equations of state, and rates of gain or loss of mass (e.g., through winds) and energy (e.g., through radiation). Our method is based on a conservative, finite-volume, second-order accurate discretization of the equations, which we solve using an unconditionally-stable implicit scheme. We implement Anderson acceleration to speed convergence of the scheme, and show that this leads to factor of $\sim 5$ speed gains over non-accelerated methods in realistic problems\red{, though the amount of speedup is highly problem-dependent}. We have implemented our method in the new code Viscous Accretion Disk Evolution Resource (\texttt{VADER}), which is freely available for download from \url{https://bitbucket.org/krumholz/vader/} under the terms of the GNU General Public License.
\end{abstract}


\end{frontmatter}

\section{Introduction}

Accretion disks are ubiquitous in astrophysics, in fields ranging from star and planet formation to high energy astrophysics to galaxies, and an enormous amount of effort has been invested in modeling them \citep[e.g.][]{pringle81a}. One approach to constructing such models is to conduct full two- or three-dimensional simulations, and this method offers the highest fidelity to the actual physical processes taking place in disks. However, such simulations are impractically computationally expensive for phenomena that take place over very large numbers of orbital timescales, or for disks where the characteristic scales that must be resolved for the simulation to converge are vastly smaller than the disk radial extent. For example, a long-term two-dimensional simulation of a protoplanetary disk might cover several thousand orbits, but the timescale over which planets form is millions of orbits. Quasi-periodic oscillations from disks around black holes, neutron stars, and white dwarfs can take place over similarly large numbers of orbits. In galaxies, the number of orbits is relatively modest, but the characteristic size scale of gravitational instability for the coldest phase of the interstellar medium is $\sim 10^6$ times smaller than the disk radial extent. None of these problems are amenable to solution by two- or three-dimensional simulations, at least not without extensive use of sub-grid models to ease the resolution requirements.

In such cases, one-dimensional simulations in which the disk is treated as vertically thin and axisymmetric are a standard modeling tool. The general approach in such simulations is to approximate the \red{turbulence} responsible for transporting mass and angular momentum through the disk as a viscosity, and to develop an analytic or semi-analytic model for this transport mechanism. Cast in this form, the evolution of a disk is described by a pair of one-dimensional parabolic partial differential equations for the transport of mass and energy; the form of these equations is analogous to a diffusion equation in cylindrical coordinates. Depending on the nature of the problem, these equations may have source or sink terms, may have a wide range of boundary conditions, and may have multiple sources of non-linearity.

Thus far in the astrophysical community most viscous disk evolution codes have been single-purpose, intended for particular physical regimes and modeling physical processes relevant to that regime. Thus for example there are codes intended for protoplanetary disks that include models for accretion onto the disk during ongoing collapse \citep[e.g.][]{hueso05a, visser10a, lyra10a, horn12a, benz14a}, galaxy disk codes containing prescriptions for star formation \citep[e.g.][]{forbes12a, forbes14a}, and codes for simulating accretion onto compact objects that include models for magnetically-dominated coronae and have equations of state that include the radiation pressure-dominated regime \citep[e.g.][]{liu02a, mayer07a, cambier13a}. While these codes are specialized to their particular problems, they are often solving very similar systems of equations, and thus there is a great deal of replication of effort in every community developing its own code.

This is particularly true because most of the codes are not open source, and for the most part the authors have not published detailed descriptions of their methodologies, forcing others to invent or re-discover their own. The sole exceptions of which we are aware \red{are the} \texttt{GIDGET} \red{code for simulating galaxy disk evolution} \citep{forbes12a} \red{and the $\alpha$-disk code published by \citet{lyra10a} and \citet{horn12a}, based on the \texttt{PENCIL} code \citep{brandenburg02a}. Neither \texttt{GIDGET} nor Lyra's code are suited for general use. For example, neither allows a wide range of viscosities, equations of state, and rotation curves.} \red{Ironically, this situation is in sharp contrast to the situation for two-dimensional disk simulations, where there are a number of open source codes that include viscosity and various other physical processes. These include \texttt{ZEUS-2D} \citep{stone92a, stone92b, stone92c}, \texttt{VHD} \citep{mckinney02a, mckinney13a}, and \texttt{PLUTO} \citep{mignone07a}. However, while it is possible to run all these codes in either a one-dimensional or pseudo-one-dimensional mode, they are all based on explicit schemes, which limits their ability to simulate very long time scales.}

The goal of this paper is to introduce a very general method for computing the time evolution of viscous, thin, axisymmetric disks \red{in one dimension, using a method suitable for simulating disks over many viscous evolution times at modest computational cost}. We embed this method in a code called Viscous Accretion Disk Evolution Resource (\texttt{VADER}), which we have released under the GNU General Public License. \texttt{VADER} is available for download from \url{https://bitbucket.org/krumholz/vader/}. The code is highly flexible and modular, and allows users to specify arbitrary rotation curves, equations of state, prescriptions for the viscosity, grid geometries, boundary conditions, and source terms for both mass and energy. The equations are written in conservation form, and the resulting algorithm conserves mass, momentum, \red{angular momentum,} and energy to machine precision, as is highly desirable for simulations of very long term evolution. We employ an implicit numerical method that is unconditionally stable, allows very large time steps, and is fast thanks to modern convergence acceleration techniques. \red{\texttt{VADER} is descended from \texttt{GIDGET} \citep{forbes12a, forbes14a} in a very general sense, but it is designed to be much more flexible and modular, while omitting many of the features (e.g., cosmological accretion and the dynamics of collisionless stars) that are specific to the problem of galaxy formation. It is implemented in C and Python. The present version is written for single processors, but we plan to develop a threaded version in the future once an open source, threaded tridiagonal matrix solver becomes available.}

The plan for the remainder of this paper is as follows. In Section \ref{sec:eqalgorithm} we introduce the underlying equations that \texttt{VADER} solves, and describe our algorithm for solving them. In Section \ref{sec:acctests} we present a number of tests of the code's accuracy and convergence characteristics. Section \ref{sec:performance} discusses the efficiency and performance of the algorithm. Finally we summarize in Section \ref{sec:summary}.

\section{Equations and Simulation Algorithm}
\label{sec:eqalgorithm}

\subsection{Equations}

The physical system that \texttt{VADER} models is a thin, axisymmetric disk of material in a time-steady gravitational potential. We consider such a disk centered at the origin and lying in the $z=0$ plane of a cylindrical $(r,\phi,z)$ coordinate system. The equations of continuity and total energy conservation for such a system, written in conservation form, are (e.g., equations 1 and A13 of \citealt{krumholz10c})
\begin{eqnarray}
\label{eq:masscons}
\ddt \Sigma + \frac{1}{r}\ddr \left(r v_r \Sigma\right) & = & \dot{\Sigma}_{\rm src} \\
\label{eq:encons}
\ddt E + \frac{1}{r}\ddr \left[rv_r\left(E+P\right)\right] - \frac{1}{r}\ddr \left(r \frac{v_\phi \mathcal{T}}{2\pi r^2}\right) & = & \dot{E}_{\rm src}.
\end{eqnarray}
Here $\Sigma$ is the mass surface density in the disk,
\begin{equation}
E = \Sigma\left(\frac{v_\phi^2}{2}+\psi\right) + E_{\rm int} \equiv \Sigma \psi_{\rm eff} + E_{\rm int}
\end{equation}
is the total energy per unit area, $v_\phi$ is the rotation speed as a function of radius, $\psi$ is the gravitational potential (which is related to $v_\phi$ by $\partial\psi/\partial r = v_\phi^2/r$), $\psi_{\rm eff} = \psi+v_\phi^2/2$ is the gravitational plus orbital energy per unit mass, $E_{\rm int}$ is the internal energy per unit area, $P$ is the vertically-integrated pressure ($\int_{-\infty}^{\infty} p\,dz$, where $p$ is the pressure), $v_r$ is the radial velocity, and $\mathcal{T}$ is the torque applied by a ring of material at radius $r$ to the adjacent ring at $r+dr$. The source terms $\dot{\Sigma}_{\rm src}$ and $\dot{E}_{\rm src}$ represent changes in the local mass and energy per unit area due to vertical transport of mass (e.g., accretion from above, mass loss due to winds, or transformation of gas into collisionless stars) or energy (e.g., radiative heating or cooling). \texttt{VADER} allows very general equations of state; the vertically-integrated pressure $P$ may be an arbitrary function of $r$, $\Sigma$, and $E_{\rm int}$ (but not an explicit function of time).

The torque and radial velocity are related via angular momentum conservation, which implies
\begin{equation}
\label{eq:vr}
v_r = \frac{1}{2\pi r\Sigma v_\phi (1+ \beta)} \ddr\mathcal{T},
\end{equation}
where $\beta = \partial \ln v_\phi /\partial \ln r$ is the index of the rotation curve at radius $r$. To proceed further we require a closure relation for the torque $\mathcal{T}$. For the purposes of this calculation we adopt the \citet{shakura73a} parameterization of this relation, slightly modified as proposed by \citet{shu92a}. In this parameterization, the viscosity is described by a dimensionless parameter $\alpha$ such that
\begin{equation}
\label{eq:alphatorque}
\mathcal{T} = -2\pi r^2 \alpha (1-\beta) P.
\end{equation}
Note that inclusion of the $1-\beta$ term is the modification proposed by \citet{shu92a}, and simply serves to ensure that the torque remains proportional to the local rate of shear in a disk with constant $\alpha$ but non-constant $\beta$. With this definition of $\alpha$, the kinematic viscosity is
\begin{equation}
\nu = \alpha \left(\frac{r}{v_\phi}\right)\left(\frac{P}{\Sigma}\right).
\end{equation}
The dimensionless viscosity $\alpha$ (as well as the source terms $\dot{\Sigma}_{\rm src}$ and $\dot{E}_{\rm src}$) can vary arbitrarily with position, time, $\Sigma$, and $E$.

\red{Since these equations are derived in \citet{krumholz10c}, we will not re-derive them here, but we will pause to comment on the assumptions that underlie them, and the potential limitations those assumptions imply. The system of equations is appropriate for a slowly-evolving thin disk with negligible radial pressure support. Specifically, we assume that (1) the scale height $H\ll r$ (thinness), (2) the radial velocity $v_r$ obeys $v_r \ll v_\phi$ and $\Sigma v_r^2 \ll E_{\rm int}$ (slow evolution), and (3) $\Sigma|v_\phi^2/r - \partial\psi/\partial r| \ll E_{\rm int}$ (negligible radial pressure support). \texttt{VADER} is not appropriate for disks that do not satisfy these assumptions.
}

\subsection{Spatial Discretization}

To discretize the equation in space, consider a grid of $N$ cells with edges located at positions $r_{-1/2}, r_{1/2}, r_{3/2}, \ldots r_{N-1/2}$ and centers at positions $r_0, r_1, r_2,\ldots r_{N-1}$. Most often the grid will be uniformly-spaced in either the logarithm of $r$, in which case $r_i = \sqrt{r_{i-1/2} r_{i+1/2}}$ and $\Delta \ln r_{i+1/2} = \ln (r_{i+1}/r_i)$ is constant, or in $r$ itself, so that $r_i = (r_{i-1/2}+r_{i+1/2})/2$ and $\Delta r_{i+1/2} = r_{i+1}-r_i$ is constant. However, \texttt{VADER} allows arbitrary placement of the cell edges, so long as the sequence $r_{i+1/2}$ is strictly increasing with $i$.

Let $A_i = \pi(r_{i+1/2}^2 - r_{i-1/2}^2)$ be the area of cell $i$. We will similarly denote the rotation curve and its logarithmic derivative evaluated at cell centers and edges by $v_{\phi,i}$, $\beta_i$, $v_{\phi, i+1/2}$, and $\beta_{i+1/2}$. Integrating equations (\ref{eq:masscons}) and (\ref{eq:encons}) over the area of cell $i$, and making use of the divergence theorem to evaluate terms involving the operator $(1/r)(\partial/\partial r)(r \cdot{})$ (which is simply the radial component of the divergence operator written out in cylindrical coordinates) gives
\begin{eqnarray}
\label{eq:massconsdiscrete}
\ddt \Sigma_i + \frac{F_{M,i+1/2}-F_{M,i-1/2}}{A_i} & = & \dot{\Sigma}_{{\rm src},i} \\
\ddt E_i + \frac{F_{E,i+1/2}-F_{E,i-1/2}}{A_i} +
\frac{F_{T,i+1/2}-F_{T,i-1/2}}{A_i} & = & \dot{E}_{{\rm src}, i},
\label{eq:enconsdiscrete}
\end{eqnarray}
where we have defined the surface density $\Sigma_i$ averaged over cell $i$ by
\begin{equation}
\Sigma_i = \frac{1}{A_i} \int_{A_i} \Sigma\, dA,
\end{equation}
and similarly for $E_i$, $\dot{\Sigma}_{{\rm src},i}$, and $\dot{E}_{{\rm src},i}$, and we have defined the fluxes at cell edges by
\begin{eqnarray}
\label{eq:massflux}
F_{M,i+1/2} &= & (2\pi r v_r \Sigma)_{r=r_{i+1/2}} \\
F_{E,i+1/2} & = & [2\pi r v_r (E+P)]_{r=r_{i+1/2}} \\
F_{T,i+1/2} & = & \left[2\pi r v_\phi \alpha (1-\beta) P \right]_{r=r_{i+1/2}},
\end{eqnarray}
and similarly for $i-1/2$. The first two terms above represent the advective fluxes of mass and enthalpy (including gravitational potential energy in the enthalpy), respectively, while the third term is the energy flux associated with work done by one ring on its neighbors through the viscous torque. Note that the fluxes as defined here are total fluxes with units of mass or energy per time, rather than flux densities with units of mass or energy per time per area.

For the purposes of numerical computation it is convenient to replace the energy equation with one for the pressure. We let
\begin{equation}
E_i = E_{\rm int}(r, \Sigma_i, P_i) + \Sigma_i \psi_{{\rm eff},i},
\end{equation}
where $E_{\rm int}(r,\Sigma,P)$ is the function giving the relationship between internal energy per unit area, surface density, and vertically-integrated pressure. Note that, for an ideal gas, $E_{\rm int}$ is a function of $P$ alone, and if the disk is vertically-isothermal then it is given by $E_{\rm int} = P/(\gamma-1)$, where $\gamma$ is the ratio of specific heats. However, we retain the general case \red{where $\gamma$ can be an explicit function of $\Sigma$, $P$, and $r$ (but not time)} to allow for more complex equations of state. \red{We do not, however, allow $\gamma = 1$ exactly, because in the truly isothermal case the energy equation vanishes and the character of the system to be solved changes fundamentally. We show below that one can approximate isothermal behavior simply by setting $\gamma = 1 + \epsilon$, where $\epsilon$ is a very small parameter, and that with this approach \texttt{VADER} has no difficulty recovering analytic solutions that apply to truly isothermal disks.}

Substituting this form for $E_i$ in equation (\ref{eq:enconsdiscrete}), applying the chain rule, and using equation (\ref{eq:massconsdiscrete}) to eliminate $\partial \Sigma_i/\partial t$ gives
\begin{equation}
\label{eq:prescons}
\ddt P_i + \frac{F_{P,i^+}-F_{P,i^-}}{A_i}
= (\gamma_i-1) \dot{E}_{{\rm int,src},i}.
\end{equation}
Here
\begin{equation}
\gamma \equiv 1+\left.\frac{\partial P}{\partial E_{\rm int}}\right|_{r,\Sigma},
\end{equation}
and the source term
\begin{equation}
\dot{E}_{\rm int,src} \equiv \dot{E}_{\rm src} - \left(\psi_{\rm eff} + \delta\frac{P}{\Sigma}\right) \dot{\Sigma}_{\rm src},
\end{equation}
where
\begin{equation}
\delta \equiv \frac{1}{\gamma-1} \left.\frac{\partial \ln P}{\partial \ln \Sigma}\right|_{r,E_{\rm int}}.
\end{equation}
For a vertically-isothermal ideal gas, $\delta = 0$ because $P$ does not change as $\Sigma$ is varied at fixed $E_{\rm int}$. From the definition of $\dot{E}_{\rm int,src}$ we can see that this term represents the time rate of change of the internal energy due to external forcing (e.g., radiative losses) evaluated \textit{at fixed $\Sigma$}. For example, if a disk is both cooling radiatively and losing mass to a wind, then $\dot{E}_{\rm int,src}$ includes the rate of change of the internal energy due to radiation alone, but not any change in the internal energy due to mass loss from the wind. Finally, we have defined the left and right pressure fluxes in cell $i$ by
\begin{eqnarray}
F_{P,i^-} & = & \left(\gamma_i-1\right) 
\left[F_{E,i-1/2} + F_{T,i-1/2} - \left(\psi_{{\rm eff},i} + \delta_i\frac{P_i}{\Sigma_i}\right) F_{M,i-1/2}\right] \\
F_{P,i^+} & = & \left(\gamma_i-1\right) \left[F_{E,i+1/2} + F_{T,i+1/2} - \left(\psi_{{\rm eff},i} + \delta_i\frac{P_i}{\Sigma_i}\right) F_{M,i+1/2}\right].
\end{eqnarray}
Note that in general $F_{P,i^+} \neq F_{P,(i+1)^-}$. An important point to emphasize is that equation (\ref{eq:prescons}) is derived from the already-discretized energy equation (\ref{eq:enconsdiscrete}). \red{Thus, if $\gamma$ and $\delta$ are known analytically, which is the case for any simple equation of state, then a numerical implementation of equation (\ref{eq:prescons}) will conserve energy to machine precision. If $\gamma$ and $\delta$ must be obtained numerically, then the accuracy of conservation would depend on the accuracy with which they are determined. In fact, as we discuss below, when $\gamma$ and $\delta$ are non-constant we use a scheme that is not explicitly conservative in any event, though in practical tests we find that the accuracy of energy conservation is within $\sim 1$ digit of machine precision.}

To proceed further, we now approximate the fluxes. The advective fluxes are proportional to $v_r$, which in turn depends on the partial derivative of the torque and thus of the pressure. We approximate this using centered differences evaluated at the cell edges. Even if the actual distribution of cell positions is uniform in neither linear or logarithmic radius, we require that a grid be classified as log-like or linear-like. For a linear grid we use the standard second-order accurate centered difference, while for logarithmic grids we rewrite the derivative $\partial \mathcal{T}/\partial r$ as $(1/r)\partial \mathcal{T}/\partial \ln r$ and use a centered difference to evaluate the derivative with respect to $\ln r$. This guarantees that the derivatives are second-order accurate in either $r$ or $\ln r$ for uniformly-spaced linear or logarithmic grids; for non-uniform grids the derivatives are first-order accurate. This choice gives a mass flux
\begin{equation}
F_{M,i+1/2} = -g_{i+1/2} 
\left[\alpha_{i+1} (1-\beta_{i+1}) r_{i+1}^2 P_{i+1} - \alpha_i (1-\beta_i) r_i^2 P_i\right],
\label{eq:massfluxdisc}
\end{equation}
where for convenience we have defined the factor
\begin{equation}
g_{i+1/2} = 
\frac{2\pi}{v_{\phi,i+1/2}(1+\beta_{i+1/2})} 
\left\{
\begin{array}{ll}
1/(r_{i+1/2}\Delta\ln r_{i+1/2}), & \mbox{(logarithmic grid)} \\
1/\Delta r_{i+1/2}, & \mbox{(linear grid)} \\
\end{array}
\right.
\end{equation}
for logarithmic and linear grids, respectively. Using the same strategy for the torque flux yields
\begin{equation}
F_{T,i+1/2} = \pi r_{i+1/2} v_{\phi,i+1/2} (1-\beta_{i+1/2}) \left(\alpha_{i+1} P_{i+1} + \alpha_{i} P_i\right).
\label{eq:torqueflux}
\end{equation}
Finally, the enthalpy flux is 
\begin{equation}
F_{E,i+1/2} = h_{i+1/2} F_{M,i+1/2},
\label{eq:enthflux}
\end{equation}
where $h_{i+1/2} = [(E+P)/\Sigma]_{i+1/2}$ is our estimate for the specific enthalpy (including the gravitational potential energy) at the cell edge. To estimate $h_{i+1/2}$, we divide the enthalpy into an internal part and a gravitational plus orbital part, i.e., we set
\begin{eqnarray}
h_{i+1/2} & = & \left(\frac{E_{\rm int}+P}{\Sigma}\right)_{i+1/2} + \psi_{{\rm eff},i+1/2} \\
& \equiv & h_{{\rm int},i+1/2} + \psi_{{\rm eff},i+1/2}.
\end{eqnarray}
The gravitational plus orbital part $\psi_{{\rm eff},i+1/2}$ is known exactly from the specification of the rotation curve, so no approximation is necessary for it. To estimate $h_{{\rm int},i+1/2}$, \texttt{VADER} uses a \red{first-order} upwind scheme \red{\citep{fletcher91a}} to maintain stability:
\begin{equation}
h_{{\rm int},i+1/2} = F_{M,i+1/2^+} h_{{\rm int},i+1/2,L} + F_{M,i+1/2^-} h_{{\rm int},i+1/2,R}
\end{equation}
where
\begin{eqnarray}
F_{M,i+1/2^+} & = & \max(F_{M,i+1/2},0) \\
F_{M,i+1/2^-} & = & \min(F_{M,i+1/2},0).
\end{eqnarray}
The left and right internal enthalpies $h_{{\rm int}, i+1/2,L}$ and $h_{{\rm int},i+1/2,R}$ represent the values on the left and right sides of the interface. \texttt{VADER} can set these values using either piecewise constant, slope-limited piecewise linear, or piecewise parabolic extrapolation, yielding first-order accurate, second-order accurate, and third-order accurate approximations, respectively. If $h_{{\rm int},i} = [(E_{\rm int}+P)/\Sigma]_i$ is the internal enthalpy evaluated at the cell center, then piecewise constant interpolation gives
\begin{eqnarray}
h_{{\rm int},i+1/2,L} & = & h_{{\rm int},i} \\
h_{{\rm int},i+1/2,R} & = & h_{{\rm int},i+1}.
\end{eqnarray}
For piecewise linear, we set the unlimited values to
\begin{eqnarray}
\lefteqn{
h_{{\rm int},i+1/2,L,{\rm nl}} = h_{{\rm int},i+1/2,R,{\rm nl}} =
}
\nonumber \\
& & 
\left\{
\begin{array}{ll}
\left[\ln(r_{i+1}/r_{i+1/2}) h_{{\rm int},i} + \ln(r_{i+1/2}/r_i)h_{{\rm int},i+1}\right]/\Delta \ln r_{i+1/2}
& \mbox{(log)} \\
\left[(r_{i+1}-r_{i+1/2}) h_{{\rm int},i}+(r_{i+1/2}-r_i)h_{{\rm int},i+1}\right]/\Delta r_{i+1/2}
& \mbox{(linear)}
\end{array}
\right.
\label{eq:hlinear}
\end{eqnarray}
and then compute the normalized slope
\begin{eqnarray}
S_L & = & \frac{h_{{\rm int},i+1/2,L,{\rm nl}}}{h_{{\rm int},i}} - 1 \\
S_R & = & 1 - \frac{h_{{\rm int},i+1/2,R,{\rm nl}}}{h_{{\rm int},i+1}}.
\end{eqnarray}
We then set the final interface values to
\begin{eqnarray}
h_{{\rm int},i+1/2,L} & = &
\left\{
\begin{array}{ll}
h_{{\rm int},i+1/2,L,{\rm nl}}, & |S_L| \leq \textrm{lim} \\
\left[1 + \textrm{sgn}(S_L)\right] h_{{\rm int},i}, & |S_L| > \textrm{lim} \\
\end{array}
\right.
\\
h_{{\rm int},i+1/2,R} & = &
\left\{
\begin{array}{ll}
h_{{\rm int},i+1/2,R,{\rm nl}}, & |S_R| \leq \textrm{lim} \\
\left[1 - \textrm{sgn}(S_R)\right] h_{{\rm int},i+1}, & |S_R| > \textrm{lim} \\
\end{array}
\right..
\end{eqnarray}
We adopt a fiducial value $\textrm{lim} = 0.1$ for the limiting parameter, which amounts to limiting the change in specific enthalpy to 10\% between a cell center and a cell edge. \red{This limiter is similar in spirit to the limited piecewise parabolic interpolation scheme of \citet{colella84a}, in that it attempts to maintain continuity in smooth regions while allowing discontinuities in non-smooth ones.} Note that for uniformly spaced grids, equation (\ref{eq:hlinear}) reduces to a simple average between the values of the two neighboring cells. For piecewise parabolic interpolation, \texttt{VADER} constructs the left and right states $h_{{\rm int},i+1/2,L}$ and $h_{{\rm int},i+1/2,R}$ from the cell-center values $h_{{\rm int},i}$ using the piecewise parabolic method \citep[section 1]{colella84a}. The reconstruction uses either $r$ or $\ln r$ as the spatial coordinate, depending on whether the grid is classified as linear or logarithmic. By default \texttt{VADER} uses piecewise linear reconstruction, as testing shows that this generally offers the best mix of accuracy and speed. This option is used for all the code tests presented below except where otherwise noted. 

With these approximations, the equations are fully discrete in space, and all discrete approximations are second-order accurate provided that the grid is uniform in either $r$ or $\ln r$.

\subsection{Discretization in Time and Iterative Solution Method}
\label{ssec:timediscretize}

\subsubsection{Formulation of the Discrete Equations}

Because $\mathcal{T}$ depends explicitly on $P$, the equations to be solved are parabolic and have the form of a non-linear diffusion equation with source and sink terms. To avoid a severe time step constraint and ensure stability, it is therefore desirable to use an implicit discretization. Let $\Sigma_i^{(n)}$, $P_i^{(n)}$ represent the state of the system at time $t_n$. We wish to know the state $\Sigma_i^{(n+1)}$, $P_i^{(n+1)}$ at time $t_{n+1} = t_n+\Delta t$. Let $\Theta$ be the time-centering parameter, such that $\Theta = 0$ corresponds to a forwards Euler discretization, $\Theta = 1/2$ to time-centered discretization \citep{crank96a}, and $\Theta=1$ to backwards Euler. \red{(For a general discussion of the different time centering choices, and why $\Theta=0$ is a poor choice for problems of this type, see \citealt{press92a}, chapter 19.)} The resulting system of implicit equations is
\begin{eqnarray}
\lefteqn{
\Sigma_{i}^{(n+1)} + \Theta \Delta t
\left[
\frac{F_{M,i+1/2}^{(n+1)} - F_{M,i-1/2}^{(n+1)}}{A_i}
- \dot{\Sigma}_{{\rm src}, i}^{(n+1)}
\right]
=
}
\nonumber \\
& &
\Sigma_{i}^{(n)} - (1-\Theta) \Delta t
\left[
\frac{F_{M,i+1/2}^{(n)} - F_{M,i-1/2}^{(n)}}{A_i}
- \dot{\Sigma}_{{\rm src}, i}^{(n)}
\right]
\label{eq:masseqn}
\\
\lefteqn{
P_i^{(n+1)} + \Theta \Delta t
\left\{
\frac{F_{P,i^+}^{(n+1)} - F_{P,i^-}^{(n+1)}}{A_i}
- \left[\gamma_i^{(n+1)}-1\right]\dot{E}_{{\rm int, src},i}^{(n+1)}
\right\}
=
}
\nonumber \\
& &
P_i^{(n)} -  
(1-\Theta) \Delta t
\left\{
\frac{F_{P,i^+}^{(n)} - F_{P,i^-}^{(n)}}{A_i}
- \left[\gamma_i^{(n)}-1\right]\dot{E}_{{\rm int, src},i}^{(n)}
\right\},
\label{eq:preseqn}
\end{eqnarray}
where superscript $^{(n)}$ or $^{(n+1)}$ indicates whether a term is to be evaluated using at time $t_n$ using column density and pressure $\Sigma_i^{(n)}$ and $P_i^{(n)}$, or at time $t_{n+1}$ using column density and pressure $\Sigma_i^{(n+1)}$ and $P_i^{(n+1)}$.

The new time pressure fluxes $F_{P,i^+}^{(n+1)}$ and $F_{P,i^-}^{(n+1)}$ depend on the new specific enthalpy $h_i^{(n+1)}$, which in turn depends on the new internal energy $E_{{\rm int},i}^{(n+1)}$. For a simple equation of state with constant $\gamma$ and $\delta=0$, we have $E_{{\rm int}} = P/(\gamma-1)$, and it is simple to close the system. The situation is also simple if the equation of state $P(r,\Sigma,E_{\rm int})$ can be inverted analytically to yield $E_{\rm int}(r,\Sigma,P)$. However, there is no guarantee that such an analytic inversion is possible, and performing the inversion numerically could be very computationally costly, since it must be done in every cell for every cycle of the iterative method we describe below. For example, if the total pressure contains significant contributions from both gas and radiation pressure, then finding the internal energy $E_{\rm int}$ given a surface density and pressure would require that we solve a quartic equation. It is also possible that $E_{\rm int}(r,\Sigma,P)$ might not be a single-valued function of $P$, in which case we would face the problem of deciding which of several possible roots is the relevant one.

To avoid these problems, when $\gamma$ and $\delta$ are not constant \texttt{VADER} evolves the internal energy along with the column density and pressure. The evolution of the internal energy is described by
\begin{equation}
\label{eq:eintexact}
\frac{\partial E_{\rm int}}{\partial t} = \frac{1}{\gamma-1} \frac{\partial P}{\partial t} + \delta \frac{P}{\Sigma}\frac{\partial \Sigma}{\partial t},
\end{equation}
which we discretize to
\begin{eqnarray}
\lefteqn{
E_{{\rm int},i}^{(n+1)} - E_{{\rm int},i}^{(n)} = \left[\frac{\Theta}{\gamma_i^{(n+1)}-1} + \frac{1-\Theta}{\gamma_i^{(n)}-1}\right] \left[P_i^{(n+1)}-P_i^{(n)}\right]
}
\nonumber \\
& &
{} +
\left[\Theta \delta_i^{(n+1)} \frac{P_i^{(n+1)}}{\Sigma_i^{(n+1)}} +
(1-\Theta) \delta_i^{(n)} \frac{P_i^{(n)}}{\Sigma_i^{(n)}}\right] 
\left[\Sigma_i^{(n+1)} - \Sigma_i^{(n)}\right].
\label{eq:eint}
\end{eqnarray}
This becomes our third evolution equation. By evolving the internal energy separately we sacrifice conservation of total energy to machine precision. However, the error we make is only of order the error introduced by the discretization of equation (\ref{eq:eintexact}) into equation (\ref{eq:eint}). We show below that in a practical example the resulting error in conservation is only marginally greater than machine precision.

For $\Theta=0$ the equations for the new time states are trivial, but such an update scheme is generally unstable\red{, or remains stable only if one obeys a time step constraint that varies as the square of the grid spacing, which would be prohibitively expensive in high-resolution simulations that must run for many viscous evolution times. For this reason,}
\texttt{VADER} supports both $\Theta = 1/2$ and $\Theta=1$. The former choice is second-order accurate in time and the latter is first-order accurate, and both choices are unconditionally stable. However, $\Theta=1/2$ is subject to spurious oscillations for sufficiently large time steps, and thus in some circumstances it may be preferable to use $\Theta=1$ despite its formally lower-order accuracy.

When $\Theta \neq 0$ equations (\ref{eq:masseqn}), (\ref{eq:preseqn}), and (\ref{eq:eint}) constitute a non-linear system, because the terms $F_{E,i-1/2}^{(n+1)}$ and $F_{E,i+1/2}^{(n+1)}$ that enter $F_{P,i^+}^{(n+1)}$ and $F_{P,i^-}^{(n+1)}$ involve products between $P_{i+1}^{(n+1)}$, $P_i^{(n+1)}$, and $P_{i-1}^{(n+1)}$. If the dimensionless viscosity $\alpha$, the source terms for mass $\dot{\Sigma}_{\rm src}$ or internal energy $\dot{E}_{\rm int,src}$, or the terms $\gamma$ and $\delta$ describing the equation of state, depend on $\Sigma$ or $P$, that represents a second source of non-linearity. Solution \red{when $\Theta \neq 0$} therefore requires an iterative approach.

\subsubsection{Linearization and Iteration Scheme}

There are numerous strategies available for solving systems of coupled non-linear equations, and our choice of method is dictated by a few considerations. \red{The most common and familiar methods of solving non-linear equations are Newton's method and its higher-dimensional generalization such as Newton-Raphson iteration. However, the simplest forms of these methods require that we be able to compute the partial derivatives of the right-hand sides of equations (\ref{eq:masseqn}), (\ref{eq:preseqn}), and (\ref{eq:eint}) with respect to $\Sigma_i^{(n)}$, $P_i^{(n)}$, and $E_{{\rm int},i}^{(n)}$. Unfortunately we cannot assume that these are available, because} the functional forms of $\alpha$, $\dot{\Sigma}_{\rm src}$ $\dot{E}_{\rm int,src}$, $\gamma$, and $\delta$ are not known \textit{a priori}, and even if their dependence on $P$ is known, this dependence may be tabulated, or may itself be the result of a non-trivial computation. In Section \ref{ssec:gidisk}, we present an example application where the latter is the case. Thus we cannot assume that the derivatives of these terms are known or easily computable. While methods such as Jacobian-free Newton-Krylov \citep[e.g.][]{knoll04a} might still be available, we instead prefer to cast the problem in terms of fixed point iteration, because it is possible to write the inner step of this iteration in a way that is particularly simple and computationally-cheap to evaluate. 

\red{Fixed point iteration is a strategy for solving non-linear equations by recasting them as the problem of finding the fixed point of a function \citep[section 2.2]{burden11a}. For example, the problem of finding the vector of values $\mathbf{x}_p$ that is a solution to a non-linear system of equations $\mathbf{f}(\mathbf{x}) = \mathbf{0}$ is obviously equivalent the problem of finding a fixed point of the function $\mathbf{g}(\mathbf{x}) = \mathbf{x}-\mathbf{f}(\mathbf{x})$, meaning that $\mathbf{g}(\mathbf{x}_p) = \mathbf{x}_p$. The simplest strategy for finding a fixed point is Picard iteration, whereby one guesses an initial value $\mathbf{x}_{0}$ and generates a new guess by setting $\mathbf{x}_{1} = \mathbf{g}(\mathbf{x}_{0})$. This procedure is then repeated until the sequence of values $\mathbf{x}_{k}$ converges to whatever tolerance is desired. One can show that, for a well-behaved function and a starting guess $\mathbf{x}_{0}$ sufficiently close to the fixed point, this procedure will indeed converge. Fixed point iteration is most useful when one can make a clever choice for the iterated function $\mathbf{g}(\mathbf{x})$. Note that one need not choose $\mathbf{g}(\mathbf{x}) = \mathbf{x} - \mathbf{f}(\mathbf{x})$ at every step of the iterative procedure; one only requires that, as the iteration number $k\rightarrow \infty$, $\mathbf{g}(\mathbf{x})$ approaches this value, or approaches any other function that has a fixed point when $\mathbf{f}(\mathbf{x}) = 0$. This means that we are free to replace $\mathbf{f}(\mathbf{x})$ with a linearized function $\mathbf{f}_L(\mathbf{x})$ that is computationally cheaper to evaluate than the true $\mathbf{f}(\mathbf{x})$, so long as the linearized form approaches the true $\mathbf{f}(\mathbf{x})$ as $k\rightarrow\infty$. This is the strategy we will adopt below.
}

\red{With that general discussion of fixed point iteration complete, we apply the method to our problem.} Let $\bmath{\Sigma}^{(n)}$, $\bmath{P}^{(n)}$, and $\bmath{E}_{\rm int}^{(n)}$ be vectors of column density and vertically-integrated pressure values in every cell at the start of a time step, and let $\bmath{q}^{(n)} = (\bmath{\Sigma}^{(n)}, \bmath{P}^{(n)}, \bmath{E}_{\rm int}^{(n)})$ be a combined vector describing the full state of the simulation at time $t_n$. We seek the vector of quantities $\bmath{q}^{(n+1)}$ that is the solution to equations (\ref{eq:masseqn}), (\ref{eq:preseqn}), and (\ref{eq:eint}). We can write the formal solution as
\begin{equation}
\label{eq:formalsolution}
\bmath{q}^{(n+1)} = F\left[\bmath{q}^{(n)}\right],
\end{equation}
where $F$ is an operator that takes the old state $\bmath{q}^{(n)}$ as an argument and returns the new state $\bmath{q}^{(n+1)}$.

Now consider a series of guesses $\bmath{q}^{(k,*)}$ for the true solution $\bmath{q}^{(n+1)}$. We wish to generate a sequence of iterates $\bmath{q}^{(k,*)}$ such that $\bmath{q}^{(k,*)} \rightarrow \bmath{q}^{(n+1)}$ as $k\rightarrow \infty$. To construct this sequence of iterates, consider a linearized version of $F$, which we denote $F_L$. The non-linear pressure equation (\ref{eq:preseqn}) can be written in matrix form as
\begin{equation}
\label{eq:matrixeqn1}
\textbfss{M} \bmath{P}^{(n+1)} = \bmath{b}
\end{equation}
where the right hand side vector $\bmath{b}$ has elements
\begin{eqnarray}
b_i & = & P_i^{(n)} - (1-\Theta) \Delta t
\left\{
\frac{F_{P,i^+}^{(n)} - F_{P,i^-}^{(n)}}{A_i}
-\left[\gamma_i^{(n)}-1\right]\dot{E}_{{\rm int,src},i}^{(n)}
\right\}
\nonumber \\
& & {}
+ \Theta \Delta t \left[\gamma_i^{(n+1)}-1\right]\dot{E}_{{\rm int, src},i}^{(n+1)},
\end{eqnarray}
and $\textbfss{M}$ is a tridiagonal matrix with elements
\begin{eqnarray}
M_{i,i+1} & = &
\Theta \frac{\Delta t}{A_i} \left[\gamma_i^{(n+1)}-1\right]\alpha_{i+1}^{(n+1)} 
\nonumber \\
& & \qquad
\left\{g_{i+1/2} r_{i+1}^2 (1-\beta_{i+1}) \left[\psi_{\rm eff,i}+\delta_i^{(n+1)}\frac{P_i^{(n+1)}}{\Sigma_i^{(n+1)}}-h_{i+1/2}^{(n+1)}\right]
\right.
\nonumber \\
& & \qquad\quad
\left.
{}\vphantom{\frac{P_i^{(n+1)}}{\Sigma_i^{(n+1)}}}
+ \pi r_{i+1/2} v_{\phi,i+1/2} (1-\beta_{i+1/2})\right\}
\\
M_{i,i} & = &
1 + \Theta \frac{\Delta t}{A_i} \left[\gamma_i^{(n+1)}-1\right]\alpha_{i}^{(n+1)} 
\nonumber \\
& & \qquad
\left\{
(1-\beta_i) r_i^2
\left[g_{i+1/2} \left(h_{i+1/2}^{(n+1)}-\psi_{{\rm eff},i}-\delta_i^{(n+1)}\frac{P_i^{(n+1)}}{\Sigma_i^{(n+1)}}\right) 
\right.\right.
\nonumber \\
& & 
\qquad\qquad\qquad\qquad
\left.
{}
+ g_{i-1/2} \left(h_{i-1/2}^{(n+1)}-\psi_{{\rm eff},i}-\delta_i^{(n+1)}\frac{P_i^{(n+1)}}{\Sigma_i^{(n+1)}}\right)\right] 
\nonumber \\
& &
\qquad\quad
\left.
\vphantom{\frac{P_i^{(n+1)}}{\Sigma_i^{(n+1)}}}
{} +
\pi \left[r_{i+1/2} v_{\phi,i+1/2}(1-\beta_{i+1/2}) 
\right.
\right.
\nonumber  \\
& &
\qquad\qquad\quad
\left.
\vphantom{\frac{P_i^{(n+1)}}{\Sigma_i^{(n+1)}}}
\left.
{}
- r_{i-1/2} v_{\phi,i-1/2}(1-\beta_{i-1/2})\right]
\right\}
\\
M_{i,i-1} & = &
\Theta \frac{\Delta t}{A_i}\left[\gamma_i^{(n+1)}-1\right]\alpha_{i-1}^{(n+1)}  
\nonumber \\
& & \qquad
\left\{g_{i-1/2} r_{i-1}^2 (1-\beta_{i-1}) \left[\psi_{\rm eff,i}+\delta_i^{(n+1)}\frac{P_i^{(n+1)}}{\Sigma_i^{(n+1)}}-h_{i-1/2}^{(n+1)}\right]
\right.
\nonumber \\
& & \qquad\quad
\left.
{}\vphantom{\frac{P_i^{(n+1)}}{\Sigma_i^{(n+1)}}}
- \pi r_{i-1/2} v_{\phi,i-1/2}(1-\beta_{i-1/2})\right\}
\\
M_{i,j} & = & 0, \quad \forall \, |i-j| > 1.
\end{eqnarray}
The equation is non-linear because $\textbfss{M}$ and $\bmath{b}$ both depend on the new quantities $\bmath{q}^{(n+1)}$, but we can construct a linearized version of the equation by instead solving
\begin{equation}
\label{eq:matrixeqn2}
\textbfss{M}_L \bmath{P}^{(\dagger)} = \bmath{b}_L,
\end{equation}
where $\textbfss{M}_L$ and $\bmath{b}_L$ differ from $\textbfss{M}$ and $\bmath{b}$ in that all quantities that depend on $\bmath{q}^{(n+1)}$ are replaced by identical terms evaluated with $\bmath{q}^{(k,*)}$. In other words, we evaluate all terms in the matrix and in the right hand side vector using the previous guess at the column density and vertically-integrated pressure. Equation (\ref{eq:matrixeqn2}) is linear in $\bmath{P}^{(\dagger)}$. Moreover, since $\textbfss{M}_L$ is tridiagonal, it is a particularly simple linear system, and can be solved with a number of operations that is linearly proportional to the number of computational cells.\footnote{\texttt{VADER}'s implementation uses the GNU Scientific Library implementation, which is based on Cholesky decomposition \red{\citep[chapter 2]{press92a}}. See \url{http://www.gnu.org/software/gsl/}.}

To complete the specification of the linearized operator $F_L$, we linearize equations (\ref{eq:masseqn}) and (\ref{eq:eint}) in an analogous manner, to
\begin{eqnarray}
\Sigma_{i}^{(\dagger)} & = &
\Sigma_i^{(n)} 
- (1-\Theta) \Delta t \left[\frac{F_{M,i+1/2}^{(n)} - F_{M,i-1/2}^{(n)}}{A_i} - \dot{\Sigma}_{{\rm src},i}^{(n)}\right]
\nonumber \\
& & \quad {}
- \Theta \Delta t \left[\frac{F_{M,i+1/2}^{(\dagger*)} - F_{M,i-1/2}^{(\dagger*)}}{A_i} - \dot{\Sigma}_{{\rm src},i}^{(\dagger*)}
\right],
\label{eq:masseqn1}
\\
E_{{\rm int},i}^{(\dagger)} & = & E_{{\rm int},i}^{(n)} + 
\left[\frac{\Theta}{\gamma_i^{(k,*)}-1} + \frac{1-\Theta}{\gamma_i^{(n)}-1}\right] \left[P_i^{(\dagger)}-P_i^{(n)}\right] 
\nonumber \\
& & \quad {}
+
\left[\Theta \delta_i^{(k,*)} \frac{P_i^{(\dagger)}}{\Sigma_i^{(\dagger)}} +
(1-\Theta) \delta_i^{(n)} \frac{P_i^{(n)}}{\Sigma_i^{(n)}}\right] 
\left[\Sigma_i^{(\dagger)} - \Sigma_i^{(n)}\right].
\label{eq:eint1}
\end{eqnarray}
where 
\begin{equation}
F_{M,i+1/2}^{(\dagger*)} = -g_{i+1/2} 
\left[\alpha_{i+1}^{(k,*)} (1-\beta_{i+1}) r_{i+1}^2 P_{i+1}^{(\dagger)} - \alpha_i^{(k,*)} (1-\beta_i) r_i^2 P_i^{(\dagger)}\right],
\end{equation}
and similarly for $F_{M,i-1/2}^{(\dagger*)}$. The mass source function $\dot{\Sigma}_{\rm src}^{(\dagger*)}$ is also evaluated using the last set of iterates for the column density, $\bmath{\Sigma}^{(k,*)}$, and the new pressure, $\bmath{P}^{(\dagger)}$, that results from solving equation (\ref{eq:matrixeqn2}). All the quantities on the right hand sides of equations (\ref{eq:masseqn1}) and (\ref{eq:eint1}) are known, and so they can be evaluated explicitly.

Thus given a starting state $\bmath{q}^{(n)}$ and a guess $\bmath{q}^{(k,*)}$ at the true solution, we can generate a new state
\begin{equation}
\bmath{q}^{(\dagger)} = F_L\left[\bmath{q}^{(k,*)}, \bmath{q}^{(n)}\right]
\end{equation}
by first solving equation (\ref{eq:matrixeqn2}) and then solving equations (\ref{eq:masseqn1}) and (\ref{eq:eint1}). Since the linearized operator $F_L$ reduces to $F$ if the first argument $\bmath{q}^{(k,*)} = \bmath{q}^{(n+1)}$, it is clear that if $\bmath{q}^{(k,*)}$ is a fixed point of $F_L$, i.e., if $\bmath{q}^{(\dagger)} = \bmath{q}^{(k,*)}$, then $\bmath{q}^{(k,*)}$ is also a solution to the original non-linear equation (\ref{eq:formalsolution}). Thus we have recast the problem of solving equations (\ref{eq:masseqn}), (\ref{eq:preseqn}), and (\ref{eq:eint}) to the problem of finding a fixed point for the linear operator $F_L$. Recall that, if the equation of state is simple and $\gamma$ and $\delta$ are constant, then we omit equation (\ref{eq:eint1}) and do not evolve $E_{{\rm int},i}$.

\subsubsection{Anderson Acceleration}

\red{
Our solution strategy can be improved further by using a convergence accelerator. In the Picard iteration procedure described in the previous section, one searches for the fixed point of a function $\mathbf{g}(\mathbf{x})$ by setting the next iterate equal to the value of the function evaluated on the previous one, i.e., by setting $\mathbf{x}_{k+1} = \mathbf{g}(\mathbf{x}_{k})$. However, this process achieves only linear convergence, meaning that, even when the guess $\mathbf{x}_{k}$ is close to the true solution $\mathbf{x}_p$, the rate at which the residuals $r_k = \left|\mathbf{x}_k - \mathbf{x}_{p}\right|$ diminish obeys $\lim_{k\rightarrow\infty} r_{k+1}/r_k = \mu$, with $\mu$ a real number in the range $(0,1)$ \citep[chapter 10]{burden11a}. That is, each iteration reduces the residual by a constant factor. The goal of a convergence accelerator is to increase the speed with which the residual diminishes. If $\mu = 0$, but $\lim_{k\rightarrow\infty} r_{k+1}/r_k^q = \mu$ for some $q>1$ and finite $\mu$, then the convergence is said to be super-linear, and the goal of a convergence accelerator is to achieve super-linearity. A side benefit of most convergence accelerators is to increase the radius of convergence, meaning that the iteration procedure will still converge to the solution $\mathbf{x}_p$ even for starting guesses $\mathbf{x}_0$ far enough from the true solution that convergence would not be achieved with Picard iteration.
}

\red{
We choose Anderson acceleration \citep{anderson65a, fang09a, walker11a} as our acceleration method. The central idea of this method is that, rather than setting our next iterate $\mathbf{x}_{k+1} = \mathbf{g}(\mathbf{x}_{k})$, we instead set it to a weighted average of the most recent iterate and $M$ previous ones, i.e., $\mathbf{x}_{k+1} = \sum_{j=0}^{M} \xi_j \mathbf{g}(\mathbf{x}_{k-j})$. The weighting coefficients are chosen such that, if the function $\mathbf{g}(\mathbf{x})$ were linear (i.e., it had a constant Jacobian), then the distance between $\mathbf{x}_{k+1}$ and the exact solution would be minimized. This condition can be expressed as a linear equation for the coefficients $\xi_j$, which can be solved using standard matrix methods. In choosing coefficients $\xi_j$ to give the exact solution if the underlying problem were linear, Anderson acceleration functions somewhat like Newton's method, which can jump directly to the solution in a single iteration when applied to a linear problem.\footnote{\red{Formally, one can show that} Anderson acceleration with $M\rightarrow \infty$ is essentially equivalent to the \red{generalized minimum residual} (GMRES, \red{\citealt{saad86a}}) method \red{for solving non-linear equations} \citep{walker11a}.} While rigorous theoretical results regarding the convergence rate of Anderson acceleration at finite $M$ have not been derived, practical tests shows that the method achieves convergence rates competitive with other Newton-like methods, which are generally super-linear \citep{calef13a, willert14a}.
}

\red{In the context of our problem, the Picard iteration method amounts to setting} $\bmath{q}^{(0,*)} = \bmath{q}^{(n)}$, and \red{then setting} all subsequent iterates via
\begin{equation}
\bmath{q}^{(k+1,*)} = \bmath{q}^{(\dagger)} = F_L\left[\bmath{q}^{(k,*)}, \bmath{q}^{(n)}\right].
\end{equation}
\red{As noted above, this strategy converges only linearly, and has a} fairly small radius of convergence, which translates into a fairly restrictive value on the time step $\Delta t$. \red{To use Anderson acceleration}, rather than setting the next iterate $\bmath{q}^{(k+1,*)}$ equal to $F_L\left[\bmath{q}^{(k,*)}, \bmath{q}^{(n)}\right]$, we set it to
\begin{equation}
\bmath{q}^{(k+1,*)} = \sum_{j=0}^{M_k} \xi_j F_L\left[\bmath{q}^{(k-j,*)}, \bmath{q}^{(n)}\right],
\end{equation}
where $M_k = \min(k, M)$, and the weight coefficients $\xi_j$ are determined by minimizing the quantity
\begin{equation}
\chi^2 = \sum_i \sum_{j=0}^{M_k} \left(\xi_j R_{ij}\right)^2
\end{equation}
subject to the constraint $\sum_j \xi_j = 1$, where
\begin{equation}
R_{ij} \equiv \frac{F_L\left[q_i^{(k-j,*)}, q_i^{(n)}\right] - q_i^{(k-j,*)}}{F_L\left[q_i^{(k-j,*)}, q_i^{(n)}\right]}
\end{equation}
is the vector of normalized residuals in every cell after iteration $k-j$. Formally, this problem is equivalent to a constrained linear least squares minimization of the overdetermined system
\begin{equation}
\textbfss{R} \bmath{\xi} = \bmath{0},
\end{equation}
where $\bmath{\xi}$ is the vector of $\xi_j$ values, $\bmath{0}$ is the 0 vector, and the constraint equation is $\bmath{1}\cdot\bmath{\xi} = 0$, where $\bmath{1}$ is the identity vector. This problem can be solved by a number of standard techniques. Our implementation in \texttt{VADER} uses QR decomposition \red{\citep[chapter 2]{press92a}}.\footnote{At present our implementation is not optimal in that we perform a QR decomposition of the residual matrix in every iteration. A faster approach would be to perform the full decomposition only during the first iteration, and then use QR factor updating techniques to recompute the QR factors directly as rows are successively added to and deleted from the residual matrix \citep{daniel76a, reichel90a}. However, at present no open source implementation of the necessary techniques is available, and constructing one is a non-trivial code development task. Since the cost of the QR decomposition is only significant in problems where the viscosity and source terms are trivial to compute, and these models are generally very quick to run in any event, we have not implemented QR factor updating at this time. For more discussion of code performance, see Section \ref{sec:performance}.}

Regardless of whether we use Picard iteration or Anderson acceleration to generate the sequence of iterates, we repeat the calculation until the residual satisfies
\begin{equation}
\label{eq:tol}
\max\left|\bmath{R}_{i,0}\right| < \mbox{tol},
\end{equation}
where the maximum is over all cells and all quantities in those cells after the most recent iteration, and tol is a pre-specified tolerance; \texttt{VADER} defaults to using $\mbox{tol} = 10^{-6}$, and we adopt this value for all the tests presented below unless noted otherwise.

\subsection{Boundary Conditions}
\label{ssec:boundary}

For the system formed by the equations of mass conservation (equation \ref{eq:massconsdiscrete}) and pressure (equation \ref{eq:prescons}), we require as boundary conditions the values of the mass flux $F_M$ and the pressure flux $F_P$ (which depends on $F_M$ and the torque and enthalpy fluxes $F_T$ and $F_E$) at the grid edge. \red{Intuitively, we can think of these two conditions as specifying the rate at which mass enters or exits the computational domain, and the rate at which energy enters or exits the computational domain.}

\red{First consider the boundary condition on the mass flux. The rate at which mass enters or exits the domain, $F_M$, can be specified in a few ways. First, one may specify this quantity directly. However, it is often more convenient to describe the mass flux in terms of the torque. Since these two are linked via angular momentum conservation (equation \ref{eq:vr}), specifying the torque, or its gradient at the domain boundary, is sufficient to specify the mass flux, and vice versa.} Thus for one of our boundary conditions we specify one of three equivalent quantities: the mass flux across the grid edge, the torque flux \red{$F_T$} across the grid edge (i.e., the rate at which the applied torque does work on the first computational zone), or the torque in a ghost zone adjacent to the grid edge.

Formally, let $i=-1$ and $i=N$ denote the indices of the ghost cells at the inner and outer boundaries of our computational grid. The quantities that we require in our update algorithm are $\alpha_{-1} P_{-1}$ and $\alpha_N P_N$, since it is the combination $\alpha P$ that appears in the definitions of the mass, torque, and enthalpy fluxes. Without loss of generality we can set $\alpha_{-1} = \alpha_0$ and $\alpha_{N} = \alpha_{N-1}$, since only the combination $\alpha P$ matters. With this choice, we set $P_{-1}$ and $P_{N}$ as follows:
\begin{enumerate}
\item  If the mass flux $F_{M,-1/2}$ or $F_{M,N-1/2}$ across the grid edge is specified, then from equation (\ref{eq:massfluxdisc}) we have
\begin{eqnarray}
\label{eq:bndfluxlower}
P_{-1} & = & \frac{(1-\beta_0)r_{0}^2}{(1-\beta_{-1})r_{-1}^2} P_0 + \frac{F_{M,-1/2}}{g_{-1/2} \alpha_{-1} (1-\beta_{-1})r_{-1}^2} \\
P_{N} & = & \frac{(1-\beta_{N-1})r_{N-1}^2}{(1-\beta_{N})r_{N}^2} P_{N-1} - \frac{F_{M,N-1/2}}{g_{N-1/2}\alpha_N (1-\beta_N)r_{N}^2}.
\end{eqnarray}
where $r_{-1} = e^{-\Delta \ln r} r_0$ for a logarithmic grid, or $r_{-1} = r_0 - \Delta r$ for a linear one, and similarly for $r_N$.
\item If the torque flux $F_{T,-1/2}$ or $F_{T,N-1/2}$ is specified, then from equation (\ref{eq:torqueflux}) we have
\begin{eqnarray}
P_{-1} & = & -P_0 + \frac{F_{T,-1/2}}{\pi r_{-1/2} v_{\phi,-1/2} (1 - \beta_{-1/2}) \alpha_{-1}} \\
P_N & = & -P_{N-1} + \frac{F_{T,N-1/2}}{\pi r_{N-1/2} v_{\phi,N-1/2} (1-\beta_{N-1/2}) \alpha_{N}}.
\label{eq:bndtorquefluxupper}
\end{eqnarray}
\item If the torque $\mathcal{T}_{-1}$ or $\mathcal{T}_N$ in the first ghost zone is specified, then from equation (\ref{eq:alphatorque}) we have
\begin{eqnarray}
P_{-1} & = & -\frac{\mathcal{T}_{-1}}{2\pi r_{-1}^2 (1-\beta_{-1})\alpha_{-1}} \\
P_{N} & = & -\frac{\mathcal{T}_{N}}{2\pi r_{N}^2 (1-\beta_{N})\alpha_N}.
\label{eq:bndtorqueupper}
\end{eqnarray}
\end{enumerate}
It should be noted that the inner and outer boundary conditions need not be specified in the same way, e.g., one can specify the  \red{mass flux indirectly by giving the} torque at the inner boundary, and \red{set} the mass flux \red{directly} at the outer boundary.

\red{Now consider the second boundary condition we require. Intuitively this is the rate of energy transport into and out of the domain, but since we are working in terms of a pressure equation, it more convenient to think in terms of the flux of enthalpy $F_E$ into or out of the domain.} From equation (\ref{eq:enthflux}), this depends on both the mass flux, and hence $\alpha P$, and on the specific internal enthalpy $h_{\rm int}$.\footnote{The enthalpy flux $F_E$ also depends on the effective gravitational potential $\psi_{{\rm eff},i}$ across the grid boundary, but we will assume that this is known from the specification of the rotation curve.} Thus we specify the final part of the boundary conditions by setting $h_{{\rm int},-1}$ and $h_{{\rm int},N}$, the values of internal enthalpy in the ghost cells. \red{Intuitively, if we know the mass flux into the domain from the first boundary condition, and we have now specified the enthalpy of the material that is being advected in, we can compute the rate at which advection is bringing energy into the computational domain. The enthalpies we require} can either be specified directly, or specified in terms of the gradient of enthalpy across the grid boundary. Note that if mass is always flowing off the grid at either the inner or outer boundary, and if one adopts piecewise constant interpolation, then the boundary value chosen for $h_{\rm int}$ will not affect the solution. \red{Also note that, because we describe the boundary condition by setting $h_{{\rm int},-1}$ rather than by setting $F_{E,-1/2}$ directly, the boundary specification is independent of the method chosen to reconstruct the enthalpies at cell edges.}

A final subtlety in choosing boundary conditions is that these conditions must be applied to both the old and new times, i.e., they must apply to both $\bmath{P}^{(n)}$ and $\bmath{P}^{(n+1)}$. The iterative solver will ensure that this holds to the accuracy of the iterative solve regardless, but by assigning appropriate values to the parts of the right hand side vector $\bmath{b}$ and matrix $\textbfss{M}$ that affect the ghost cells (i.e., elements $M_{-1,-1}$, $M_{-1,0}$, $M_{N,N-1}$, $M_{N,N}$, $b_{-1}$ and $b_N$) one can enforce the boundary conditions \red{to machine precision} at both the old and new times, and in the process speed convergence of the iterative solution step. The boundary conditions take the form of a relationship between the ghost cells and the adjacent real cells,\footnote{In general it is also possible to have boundary conditions such that values in the ghost cells depend on the values in real cells that are not immediately adjacent to the ghost zones. In this case the best approach is to treat the boundary condition as simply a specified value of $P_{-1}$ or $P_N$, without attempting to enforce this relationship at both the old and new times by setting elements of $\textbfss{M}$. Including these constraints in $\textbfss{M}$ would render $\textbfss{M}$ no longer tridiagonal. Since this would increase the cost of solving the linear system considerably, it is more efficient to include only the relationship between the ghost cells and their nearest neighbors in $\textbfss{M}$, and enforce any other relationships between the ghost and real cells by iterating the system to convergence.} as specified by equations (\ref{eq:bndfluxlower}) -- (\ref{eq:bndtorqueupper}), which can all be written in the form
\begin{eqnarray}
P_{-1} & = & q_{-1} P_0 + p_{-1}\\
P_{N} & = & q_N P_{N-1} + p_N,
\end{eqnarray}
with the constants $q$ and $p$ depending on how the boundary condition is specified. For a given choice of $q$ and $p$, the matrix and vector elements required to enforce the boundary conditions are
\begin{eqnarray}
M_{-1,-1} = M_{N,N} & = & 1 \\
M_{-1,0} & = & -q_{-1} \\
M_{N,N-1} & = & -q_N \\
b_{-1} & = & p_{-1} \\
b_N & = & p_N.
\end{eqnarray}

\subsection{Time Step Control}
\label{ssec:timestep}

The final piece required to complete the update algorithm is a recipe to choose the time step $\Delta t$. Using either $\Theta=1/2$ or 1 the system is unconditionally stable, but accuracy requires that the time step be chosen so that it is not too large. Moreover, if the time step is too large, then calculations with $\Theta=1/2$ are subject to spurious oscillations. \texttt{VADER} controls the time step based on the rate of change computed in the previous time step. If $\bmath{q}^{(n-1)}$ was the state of the system at time $t_{n-1}$, $\bmath{q}^{(n)}$ is the state at time $t_n$, and $\Delta t_{n-1}$ was the previous time step, then the next time step $\Delta t_n = t_{n+1} - t_n$ is determined by
\begin{equation}
\label{eq:timestep}
\Delta t_n = C 
\min_{i=0, 1,\ldots N-1}\left(\left|\frac{q_i^{(n-1)}}{q_i^{(n)} - q_i^{(n-1)}}\right|\right) \Delta t_{n-1}.
\end{equation}
where the minimum is over all cells, and $C$ is a dimensionless constant for which we choose a fiducial value of 0.1. To initialize the calculation, \texttt{VADER} takes a fake time step of size $10^{-4} r_{-1/2}/v_{\phi,-1/2}$, during which the code computes a new state and uses it to determine the time step, but does not actually update the state.

While the update algorithm is stable regardless of the choice of time step, the same is not necessarily true of the iterative solution procedure described in Section \ref{ssec:timediscretize}, which may not converge if $\Delta t$ is too large. The maximum time step for which convergence occurs will in general depend on the boundary conditions and the functional forms of $\alpha$, $\gamma$, $\delta$, $\dot{\Sigma}_{\rm src}$, and $\dot{E}_{\rm src}$. To ensure the stability of the overall calculation, \texttt{VADER} allows a user-specified maximum number of iterations used in the implicit solve. If convergence is not reached within this number of iterations, or if the iteration diverges entirely, \texttt{VADER} stops iterating and tries to advance again with a factor of 2 smaller time step. This ratcheting is applied recursively as necessary until convergence is achieved.

\subsection{Implementation Notes}

We have implemented the algorithm described above in the Viscous Accretion Disk Evolution Resource (\texttt{VADER}), and made the code publicly available from \url{https://bitbucket.org/krumholz/vader/} under the terms of the GNU Public License. The core code is written in c, with templates provided for user-supplied implementations of functions characterizing the inner and outer boundary conditions, the dimensionless viscosity, the equation of state terms, and the rates of mass and energy gain or loss per unit area. Each of these quantities can also be specified simply by a numerical value, so that users who do not wish to use some particular aspect of the code (e.g.~the general equation of state) need not implement functions describing it. \texttt{VADER} also provides routines to construct rotation curves suitable for computational use from tabulated $v_\phi(r)$ values; see \ref{app:tabrotation} for details.

In addition to the core C routines, \texttt{VADER} includes a set of Python wrapper routines that allow simulations to be run from a Python program. The Python wrappers provide high-level routines for processing simulation input and outputs, and controlling execution.

The \texttt{VADER} repository includes an extensive User's Guide that fully documents all functions.

\section{Accuracy and Convergence Tests}
\label{sec:acctests}

In this section we present tests of the accuracy and convergence characteristics of the code, comparing against a variety of analytic solutions, and demonstrating a number of the code's capabilities, including linear and logarithmic grids, a range of rotation curves, and arbitrary functional forms for energy gain and loss, boundary conditions, and viscosity. The code required to run all the tests described in this section is included in the \texttt{VADER} bitbucket repository.

One subtlety that occurs in all these comparisons is treatment of the boundary conditions. The true nature of the boundary layers of accretion disks is subject to significant theoretical uncertainty \citep[e.g.,][]{papaloizou86a, popham92a, popham93a}, but \red{most} known analytic solutions for viscous disks generally prescribe boundary conditions by specifying that the disk extends all the way from $r=0$ to $r=\infty$, and by demanding regularity as $r\rightarrow 0$ and $r\rightarrow \infty$. \red{(For an exception see \citet{tanaka11a}, who derives analytic solutions for disks with power law viscosities and finite inner radii.)} For obvious reasons in a numerical computation (and in nature) it is necessary to truncate the disk at finite values of $r$, and thus some care is required to match the boundary conditions used in the numerical computation to those assumed in the analytic solutions. For most reasonable choices of boundary condition, the choice will affect the results only in the vicinity of the boundaries, but for the purpose of making quantitative comparisons one must be careful to match boundary conditions between the analytic and numerical solutions to ensure that differences caused by different boundary conditions do not dominate the error budget.

\subsection{Self-Similar Disks}
\label{ssec:selfsim}

The first test is a comparison to the similarity solution derived by \citet{lynden-bell74a}. The solution is for Keplerian rotation ($\beta=-1/2$) and a kinematic viscosity $\nu$ that varies with radius as $\nu = \nu_0 (r/R_0)$, giving
\begin{equation}
\alpha = \left(\frac{\nu_0 v_\phi}{R_0}\right)\frac{\Sigma}{P}.
\end{equation}
Note that this choice of $\alpha$ renders the torque a function of $\Sigma$ only, so that the transports of mass and energy are decoupled and the rates of transport do not depend on $P$. With this rotation curve and viscosity, the evolution equations admit a similarity solution
\begin{equation}
\label{eq:selfsimanalyt}
\Sigma = \Sigma_0 \frac{e^{-x/T}}{x T^{3/2}},
\end{equation}
where $\Sigma_0 = \dot{M}_0/(3\pi \nu_0)$, $x = r/R_0$, $T = t/t_s$, $t_s = R_0^2/3\nu_0$, and $\dot{M}_0$ is the mass accretion rate reaching the origin at time $T=1$. To check \texttt{VADER}'s ability to reproduce this solution, we simulate a computational domain from $r/R_0 = 0.1 - 20$, using a uniform logarithmic grid. We initialize the grid using the analytic solution at time $T = 1$, and specify the boundary conditions by setting the torque in the ghost zones equal to the values for the similarity solution,
\begin{equation}
\mathcal{T} = -\dot{M}_0 v_\phi R_0 \frac{x}{T^{3/2}}e^{-x/T}.
\end{equation}
The equation of state is a simple one, with $\gamma = 1 + 10^{-6}$ and $\delta=0$, and the initial value of $P/\Sigma$ is set to a constant. The value of $\gamma$ is chosen so that $P/\Sigma$ remains nearly constant.

\begin{figure}
\begin{center}
\includegraphics[width=1.0\textwidth]{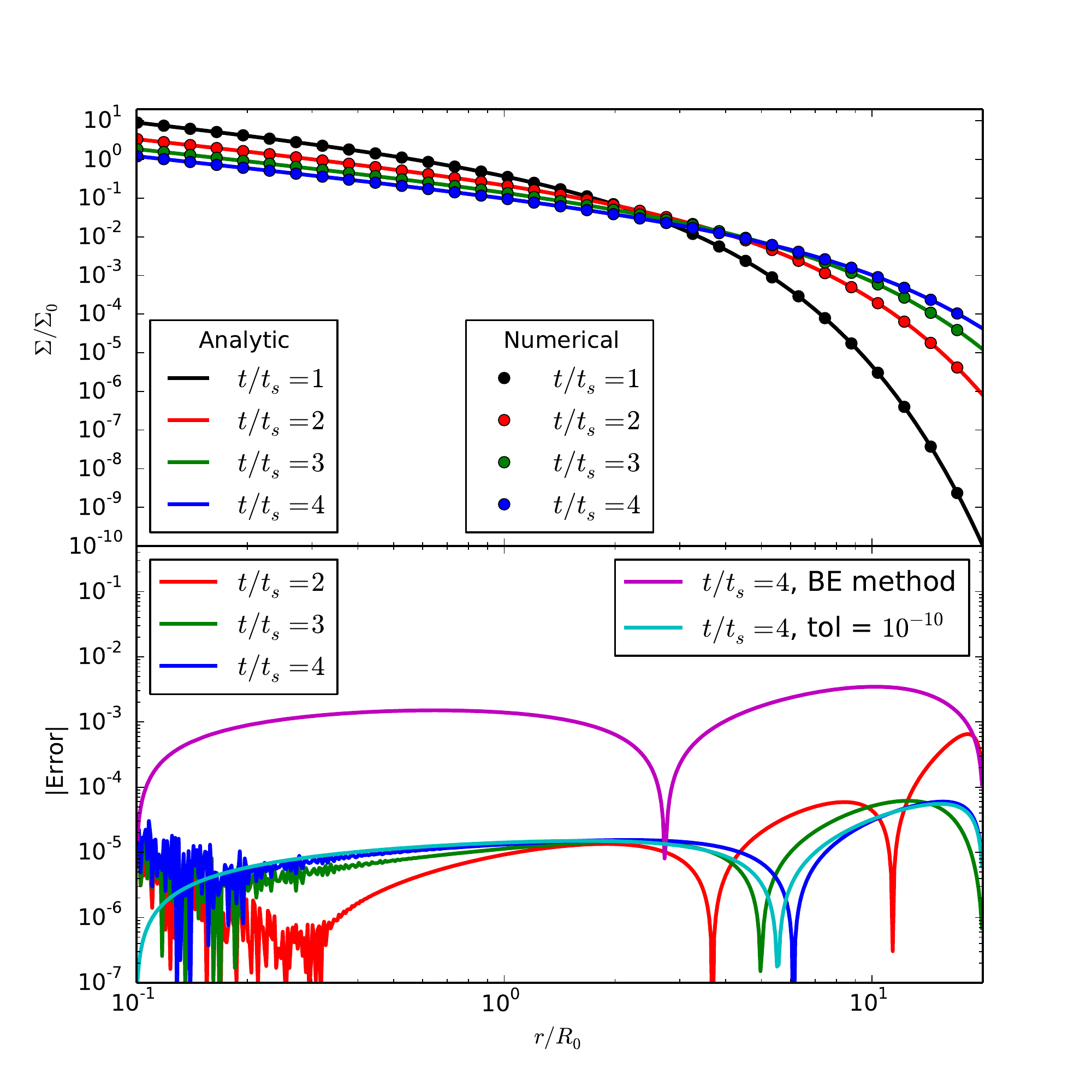}
\end{center}
\caption{
\label{fig:selfsim1}
Comparison between the analytic solution for the self-similar evolution of a viscous disk (equation \ref{eq:selfsimanalyt}) and a \texttt{VADER} simulation. The upper panel shows the normalized gas surface density $\Sigma/\Sigma_0$ versus normalized radius $r/R_0$ at several times for the analytic solution (solid lines) and the simulation (circles; only every eigth data point plotted to avoid confusion). The values plotted for $t/t_s=1$ are the initial conditions in the simulation. The lower panel shows the absolute value of the error, defined per equation (\ref{eq:selfsimerr}). The red, green, and blue lines show the same calculation as in the upper panel. The magenta line shows the result at $t/t_s = 4$ using a backwards Euler method instead of a Crank-Nicolson one, while the cyan line shows a result using a Crank-Nicolson method but with a tighter error tolerance of $10^{-10}$ instead of $10^{-6}$ in the iterative solution step.
}
\end{figure}

Figure \ref{fig:selfsim1} shows a comparison between the analytic solution and the results of a \texttt{VADER} simulation performed with a resolution of 512 cells, piecewise-linear reconstruction, a tolerance $C=0.1$ for the maximum change per time step, and Crank-Nicolson time centering. \red{We will discuss performance in more detail below, we note that this entire calculation runs in $2-3$ seconds on a single CPU.} As shown by the figure, the absolute value of the error, defined as
\begin{equation}
\label{eq:selfsimerr}
\mbox{Error} = \frac{\Sigma_{\rm numerical} - \Sigma_{\rm analytic}}{\Sigma_{\rm analytic}},
\end{equation}
is generally of order $10^{-6} - 10^{-5}$, with a maximum of about $5\times 10^{-4}$ at time $t/t_s=2$. Note that a small amount of grid-scale oscillation is visible in the error at small radii, although the overall error is still $\sim 10^{-6}$. Oscillations of this sort are a common feature of calculations using Crank-Nicolson time centering, and can be eliminated by use of the backwards Euler method, at the price of lower overall accuracy. The magenta line in Figure \ref{fig:selfsim1} shows an otherwise identical calculation using backwards Euler. The grid noise in the Crank-Nicolson result can also be greatly reduced by using a more stringent tolerance in the iterative solve. Using $\textrm{tol} = 10^{-10}$ instead of the default value of $10^{-6}$, renders the oscillation invisibly small. This is shown by the cyan line in Figure \ref{fig:selfsim1}.

\begin{figure}
\begin{center}
\includegraphics[width=1.0\textwidth]{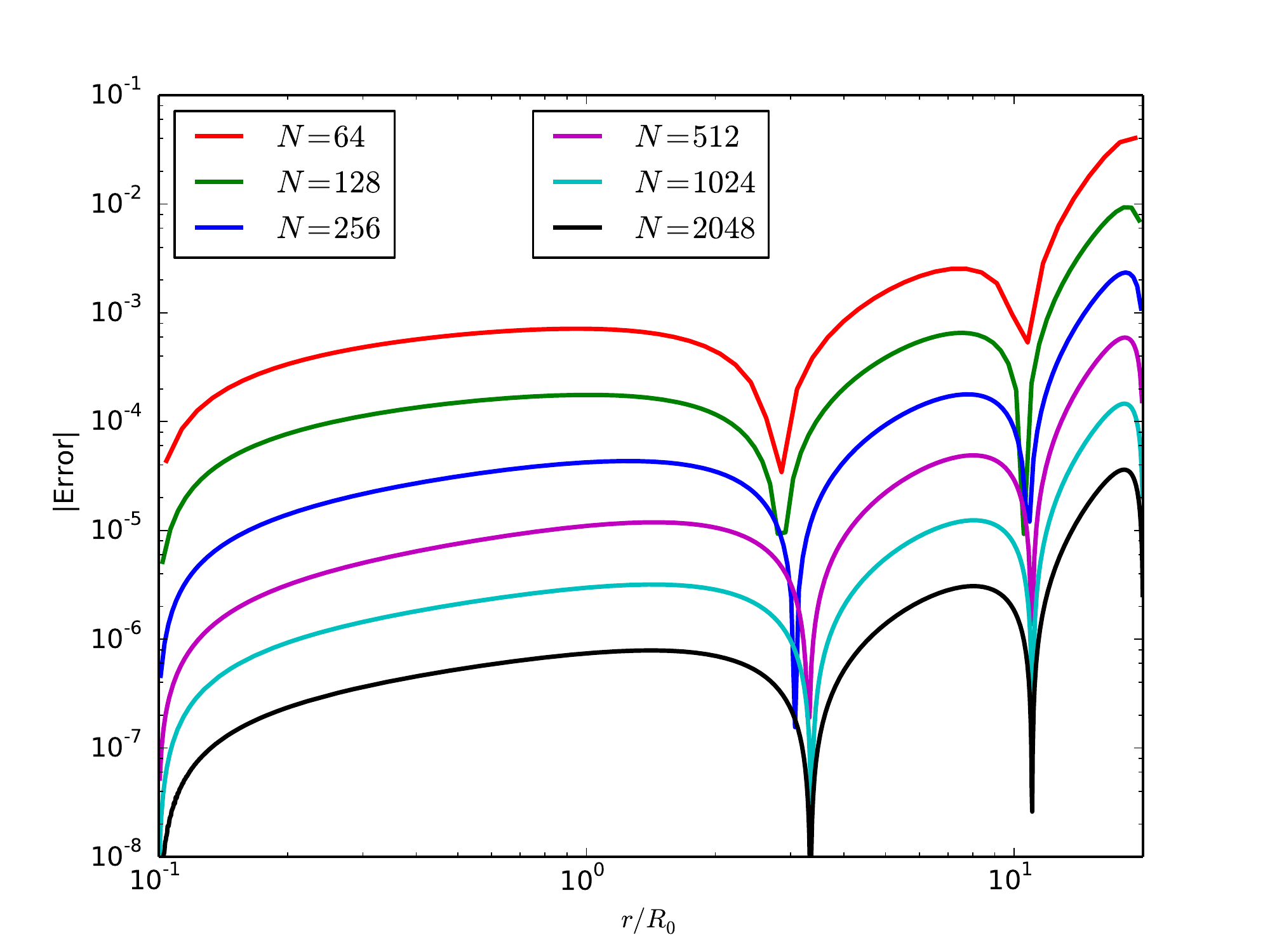}
\end{center}
\caption{
\label{fig:selfsim_resolution1}
Same as the lower panel of Figure \ref{fig:selfsim1}, but now showing the absolute value of the error versus position at time $t/t_s = 2$ for a series of computations with different numbers of cells $N$. All the calculations shown use Crank-Nicolson time centering and iterative solver tolerance $\textrm{tol} = 10^{-10}$.
}
\end{figure}

\begin{figure}
\begin{center}
\includegraphics[width=1.0\textwidth]{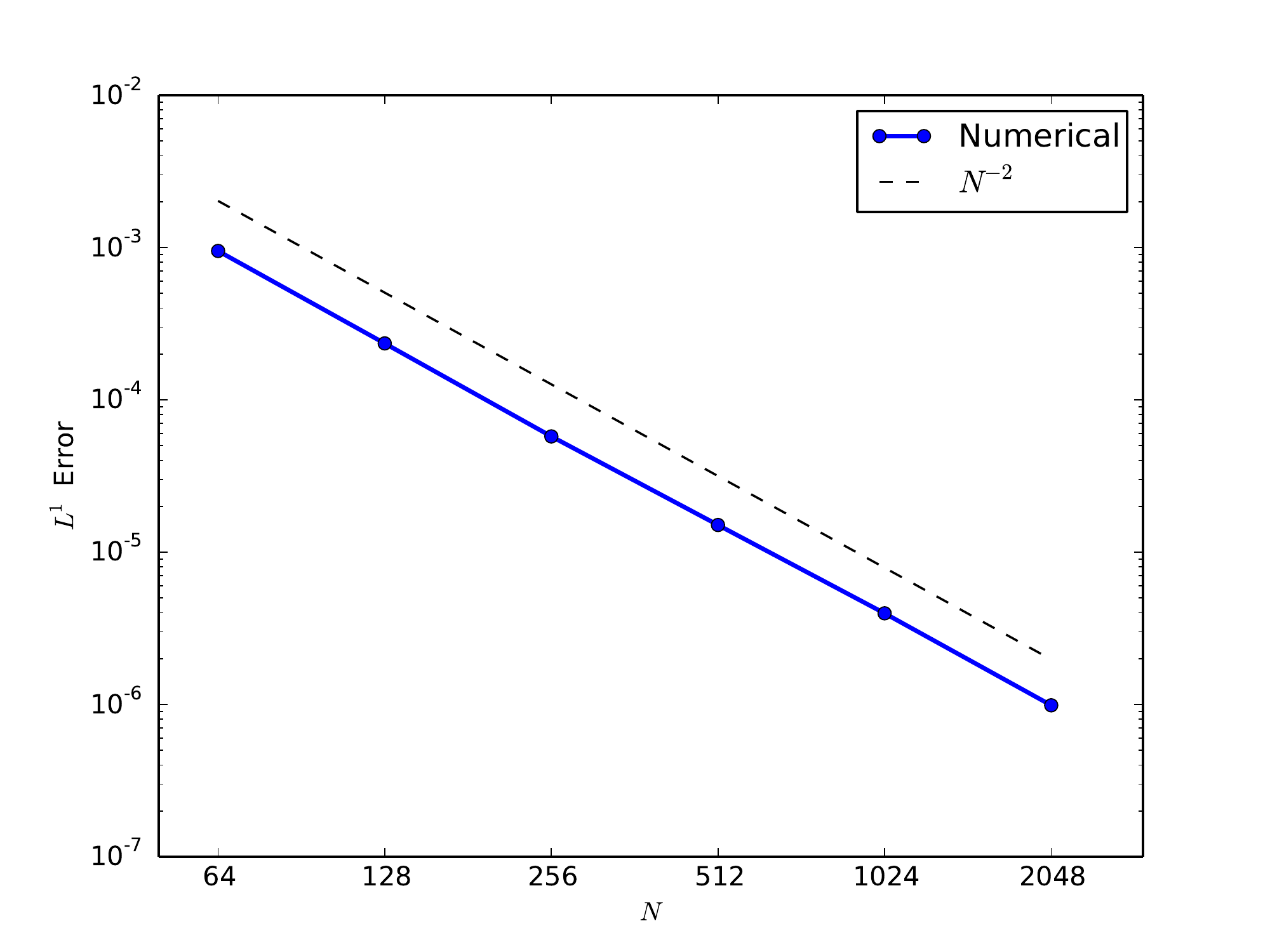}
\end{center}
\caption{
\label{fig:selfsim_resolution2}
$L^1$ error (equation \ref{eq:l1err}) versus resolution $N$ at time $t/t_s=2$ for the self-similar disk test results shown in Figure \ref{fig:selfsim_resolution1} (blue lines and points). The black dashed line shows a slope of $-2$ for comparison.
}
\end{figure}

To determine how the accuracy of the solution depends upon spatial resolution, we repeat the calculation using Crank-Nicolson time centering at a range of resolutions from $N=64$ to 2048 in factor of 2 steps. We use $\textrm{tol} = 10^{-10}$ for this test to ensure that the error is determined by the spatial resolution, and not the choice of error tolerance in the iterative solver. Figure \ref{fig:selfsim_resolution1} shows the absolute value of the error versus position at a time $t/t_s = 2$ for these calculations, and Figure \ref{fig:selfsim_resolution2} shows the $L^1$ error in the solution at this time versus resolution, where the $L^1$ (Lebesgue) norm of the error has its usual definition
\begin{equation}
\label{eq:l1err}
L^1\mbox{ error} = \frac{1}{\pi \Sigma_0 R_0^2}\sum_i A_i |\Sigma_{{\rm numerical},i} - \Sigma_{{\rm exact},i}|.
\end{equation}
As expected, the error declines with resolution, with the $L^1$ error declining as $N^{-2}$. A least-squares fit of the logarithm of the $L^1$ error versus the logarithm of resolution gives a slope of $-2.0$. This confirms that the code is second-order accurate in space, as expected. We do not perform a similar test of the time accuracy, because the actual simulation time step is not controlled by a single parameter. It depends on the time step tolerance $C$, but also on the error tolerance tol used in the iterative solver, and on the maximum number of iterations the solver is allowed to perform before giving up and trying again with a reduced time step.

\subsection{Singular Ring Test}

The second test we present is a comparison of \texttt{VADER} to the analytic solution for the evolution of an initially-singular ring of material with constant kinematic viscosity \citep{pringle81a}. Consider a disk whose initial density distribution is concentrated at a single radius $r=R_0$, such that
\begin{equation}
\Sigma(r,t=0) = m_0\frac{\delta(r - R_0)}{2\pi R_0}.
\end{equation}
The material orbits in a Keplerian potential, and has a constant kinematic viscosity $\nu$, corresponding to
\begin{equation}
\alpha = \nu \left(\frac{\Sigma}{P}\right) \left(\frac{v_\phi}{r}\right).
\label{eq:alpharing}
\end{equation}
As in the previous test problem, this choice of $\alpha$ renders the torque a function of $\Sigma$ only. With this choice of $\alpha$, and subject to the boundary conditions that $\Sigma$ remain finite and $\mathcal{T}\rightarrow 0$ as $r\rightarrow 0$,\footnote{Because the equations for $\Sigma$ and $P$ are decoupled, only two boundary conditions are required to specify the solution for $\Sigma$.} the system has the exact solution
\begin{equation}
\Sigma_{\rm exact} = \Sigma_0 \frac{1}{x^{1/4}\tau} e^{-(1+x^2)/\tau} I_{1/4}(2x/\tau),
\label{eq:ringexact}
\end{equation}
where $x = r/R_0$, $\tau= t/t_s$, $\Sigma_0 = M_0/\pi R_0^2$, and $I_{1/4}$ is the modified Bessel function of the first kind of order $1/4$. Here $t_s = r_0^2/12\nu$ is the characteristic viscous evolution time.

\begin{figure}
\begin{center}
\includegraphics[width=1.0\textwidth]{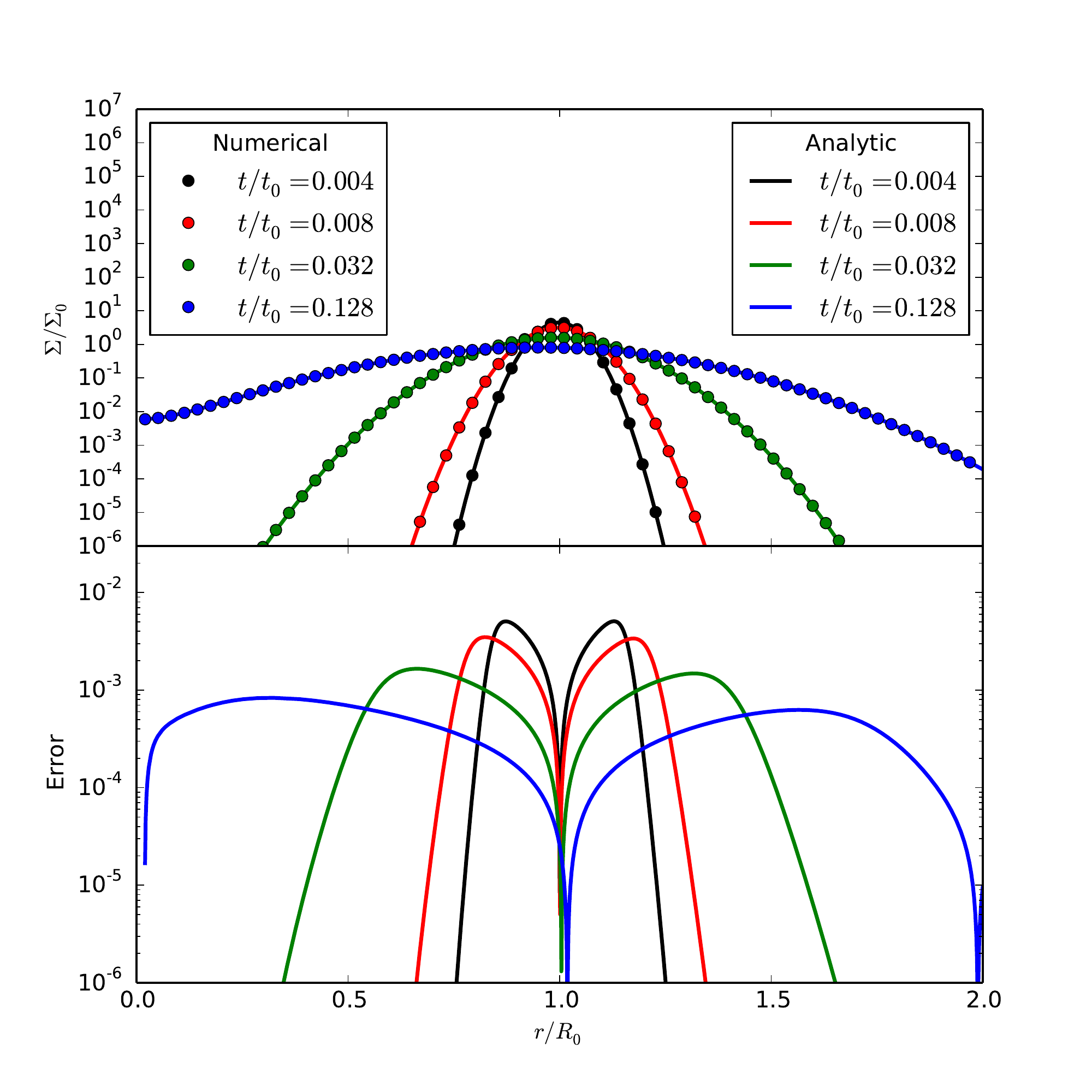}
\end{center}
\caption{
\label{fig:ring1}
Results of a simulation of an initially-singular ring test. The upper panel shows the exact analytic solution (solid lines, equation \ref{eq:ringexact}) and the numerical results produced by \texttt{VADER} (circles; only every 64th point shown, for clarity) as a function of position $r/R_0$ at times $t/t_0 = 0.004$, 0.008, 0.032, and 0.128. The lower panel shows the absolute value of the error in the numerical result, defined as in equation (\ref{eq:ringerror}).
}
\end{figure}

Due to the singular initial condition, this is a far more challenging test of the algorithm than the similarity solution discussed in the previous section. To test \texttt{VADER}'s performance on this problem, we simulate a disk on a uniform linear grid extending from $0.1R_0 - 2R_0$, initialized such that the cell containing the radius $R_0$ has a column density $\Sigma = \Sigma_{\rm init} \equiv M_0/A$, where $A$ is the area of the cell. All other cells have column densities $\Sigma = \Sigma_{\rm init}/\chi$ with $\chi=10^{10}$. The initial vertically-integrated pressure is set so that $P/\Sigma$ is constant, and the simulation uses an equation of state with $\gamma = 1+10^{-6}$, $\delta=0$, but neither of these choices affects the evolution of the gas surface density. We set the torques at the inner and outer boundaries equal to their analytic values (subject to the density floor),
\begin{equation}
\mathcal{T} = -3\pi r \nu v_\phi \Sigma,
\end{equation}
where $\Sigma = \max(\Sigma_{\rm exact}, \Sigma_{\rm init}/\chi)$. Note that the boundary torque is therefore time-dependent. Figure \ref{fig:ring1} shows the results of the test for a simulation with 4096 cells, \red{which required $20-30$ seconds of run time to complete}. As the plot shows, \texttt{VADER} reproduces the analytic result very accurately. The error, defined to account for the effects of the density floor as
\begin{equation}
\label{eq:ringerror}
\mbox{Error} = \frac{\Sigma_{\rm numerical} - \Sigma_{\rm exact} - \Sigma_{\rm init}/\chi}{\Sigma_{\rm exact} + \Sigma_{\rm init}/\chi},
\end{equation}
is of order $10^{-3}$ at very early times when the ring is poorly resolved, and drops to $\sim 10^{-4}$ or less at late times. 

\subsection{Gravitational Instability-Dominated Disks}
\label{ssec:gidisk}

The third test we present is against the analytic solution for gravitational instability-dominated disks first derived in a special case by \citet{bertin99a}, generalized by \citet{krumholz10c}, and further generalized in the time-dependent case following \citet{forbes12a} and \citet{forbes14a}. This test demonstrates \texttt{VADER}'s ability to handle a much more complex problem than the previous two tests. In this problem the rotation curve is not Keplerian, radiative cooling is an integral part of the problem, and the viscosity is not given by a simple analytic formula but instead is derived from an underlying set of equations to be solved at each time step.

Consider a pure gas disk (i.e., one with no stellar component) of surface density $\Sigma$, within which support against self-gravity is dominated by a highly supersonic velocity dispersion $\sigma$, such that the vertically-integrated pressure $P = \Sigma\sigma^2$, $\gamma=5/3$, and $\delta = 0$. The stability of the disk against axisymmetric gravitational instabilities is described by the \citet{toomre64a} $Q$ parameter, $Q = \kappa \sigma/\pi G \Sigma$, where $\kappa = [2(\beta+1)]^{1/2} v_\phi/r$ is the epicyclic frequency.\footnote{\red{Formally Toomre's $Q$ applies only to a disk where $\sigma$ is the thermal velocity dispersion, but a generalized $Q$ applies to gas where the non-thermal velocity dispersion is much greater than the sound speed \citep[e.g.,][]{elmegreen11a}.}} The turbulence in the disk decays following 
\begin{equation}
\dot{E}_{\rm int, src} = -\eta \Sigma \sigma^2 \frac{v_\phi}{r},
\end{equation}
where $\eta= 3/2$ corresponds to the full kinetic energy of the turbulence being lost for every crossing time of the scale height. This decay of turbulence is offset by mass accretion, which converts orbital energy into turbulent motion. \citet{krumholz10c} show that the rate of change of $Q$ is related to the torque implicitly via
\begin{equation}
\label{eq:torqueeqn}
\tau'' + h_1 \tau' + h_0 \tau = H,
\end{equation}
where $\tau = \mathcal{T}/[\dot{M}_R v_\phi(R) R]$ is the dimensionless torque, $R$ is a chosen scale radius, $v_\phi(R)$ and $\dot{M}_R$ are the rotation curve speed and initial (inward) accretion rate at that radius, and the coefficients appearing in the equation are
\begin{eqnarray}
h_0 & = & (\beta^2-1) \frac{u^2}{2 x^2 s^2} \\
h_1 & = & -\frac{5(\beta+1) x s' + 2 s (\beta+\beta^2+x \beta')}{2(\beta+1) s x} \\
H & = & \sqrt{\frac{(\beta+1)^3}{2\pi^2 \chi^2}} \left(2\pi \eta u - 3 \frac{d\ln Q}{d\ln T}\right) \frac{s u^2}{Q x}
\end{eqnarray}
where $x=r/R$, $u = v_\phi/v_\phi(R)$, $s=\sigma/v_\phi(R)$, $\chi = G \dot{M}_R/v_\phi(R)^3$, $T=t/t_{\rm orb}$, $t_{\rm orb} = 2\pi R/v_{\phi}(R)$, and primes indicate differentiation with respect to $x$.

If $d\ln Q/d\ln T = 0$ for $Q=1$, and $\beta$ is constant, then the combined system formed by equation (\ref{eq:torqueeqn}) and the equations of mass and energy conservation (\ref{eq:masscons}) and (\ref{eq:encons}) admit a steady-state solution. For a flat rotation curve, \red{i.e., one with $v_\phi$ constant and $\beta=0$,} this is
\begin{eqnarray}
\mathcal{T} & = & -r \dot{M}_R v_\phi(R) \\
\label{eq:Sigmasteady}
\Sigma & = & \frac{v_\phi}{\pi G r} \left(\frac{G\dot{M}_R}{\eta}\right)^{1/3} \\
\label{eq:sigmasteady}
\sigma & = & \frac{1}{\sqrt{2}} \left(\frac{G\dot{M}_R}{\eta}\right)^{1/3}.
\end{eqnarray}
If $d\ln Q / d\ln T \leq 0$ when $Q > 1$ (as is expected, since when $Q>1$ there should be no gravitational instability to offset the decay of turbulent motions) then this solution is an attractor, so that disks that start in different configurations will approach this configuration over a viscous transport time.

To study \texttt{VADER}'s ability to solve this problem, we perform a series of tests. In all of these simulations we obtain the viscous torque $\mathcal{T}$ and thus the dimensionless viscosity $\alpha$ required by \texttt{VADER} by solving a discretized version of equation (\ref{eq:torqueeqn}) on the grid with
\begin{equation}
\frac{d\ln Q}{d\ln T} = \frac{u}{x} \min(e^{-1/Q} - e^{-1}, 0),
\end{equation}
so that the disk returns to $Q = 1$ on a timescale comparable to the local orbital time. This particular functional form is not particularly physically motivated, and is chosen simply to ensure that $d\ln Q/d\ln T$ goes smoothly to 0 as $Q\rightarrow 1$ from above. The boundary conditions on equation (\ref{eq:torqueeqn}) are that $\tau=-x$ at the inner boundary and $\tau = -\beta - 1$ at the outer boundary, consistent with the steady state solution. In regions where $Q > 1$, we suppress gravitational instability-driven transport by reducing the torque by a factor $e^{-10(Q-1)}$. The \texttt{VADER} simulation uses an outer boundary condition with a fixed mass flux $\dot{M}$, and an inner boundary condition whereby the fixed torque is given by $\tau = -x e^{-10(Q-1)}$, where $Q$ is evaluated in the first grid zone. All simulations use a uniform logarithmic grid of 512 cells that goes from $0.01 R - R$, and piecewise constant enthalpy advection.

\begin{figure}
\begin{center}
\includegraphics[width=1.0\textwidth]{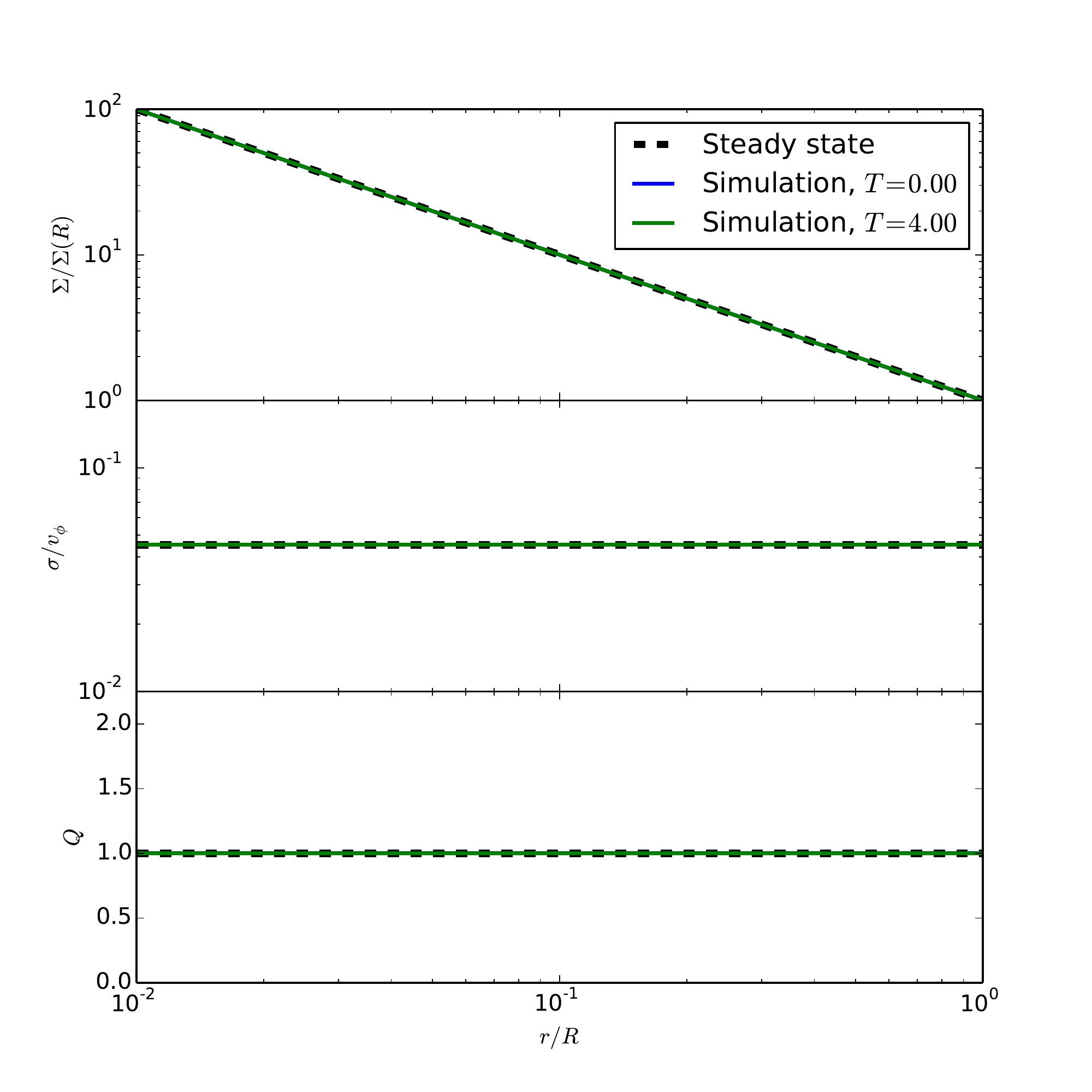}
\end{center}
\caption{
\label{fig:gidisk1}
Results of a simulation of the gravitational instability-dominated disk test. The three panels shows the gas surface density $\Sigma$ normalized to the steady-state value at $R$, the middle panel shows the velocity dispersion $\sigma$ normalized to $v_\phi$, and the bottom panel shows Toomre $Q$. The black dashed line shows the analytic steady state solution (equations \ref{eq:Sigmasteady} and \ref{eq:sigmasteady}), the blue line shows the simulation initial condition, and the green line shows the simulation after $T=4$ outer orbits. \red{Note that the blue line is completely hidden by the green line, as it should be since we are testing the ability of the code to maintain the correct analytic steady state.}
}
\end{figure}

Figure \ref{fig:gidisk1} shows the results of a simulation where the system is initialized to the analytic steady-state solution and allowed to evolve for 4 outer orbital times, corresponding to 400 orbital times at the inner edge of the disk. As the figure shows, \texttt{VADER} successfully maintains the steady state. \red{The total time required to run this computation was $20-30$ seconds.}

\begin{figure}
\begin{center}
\includegraphics[width=1.0\textwidth]{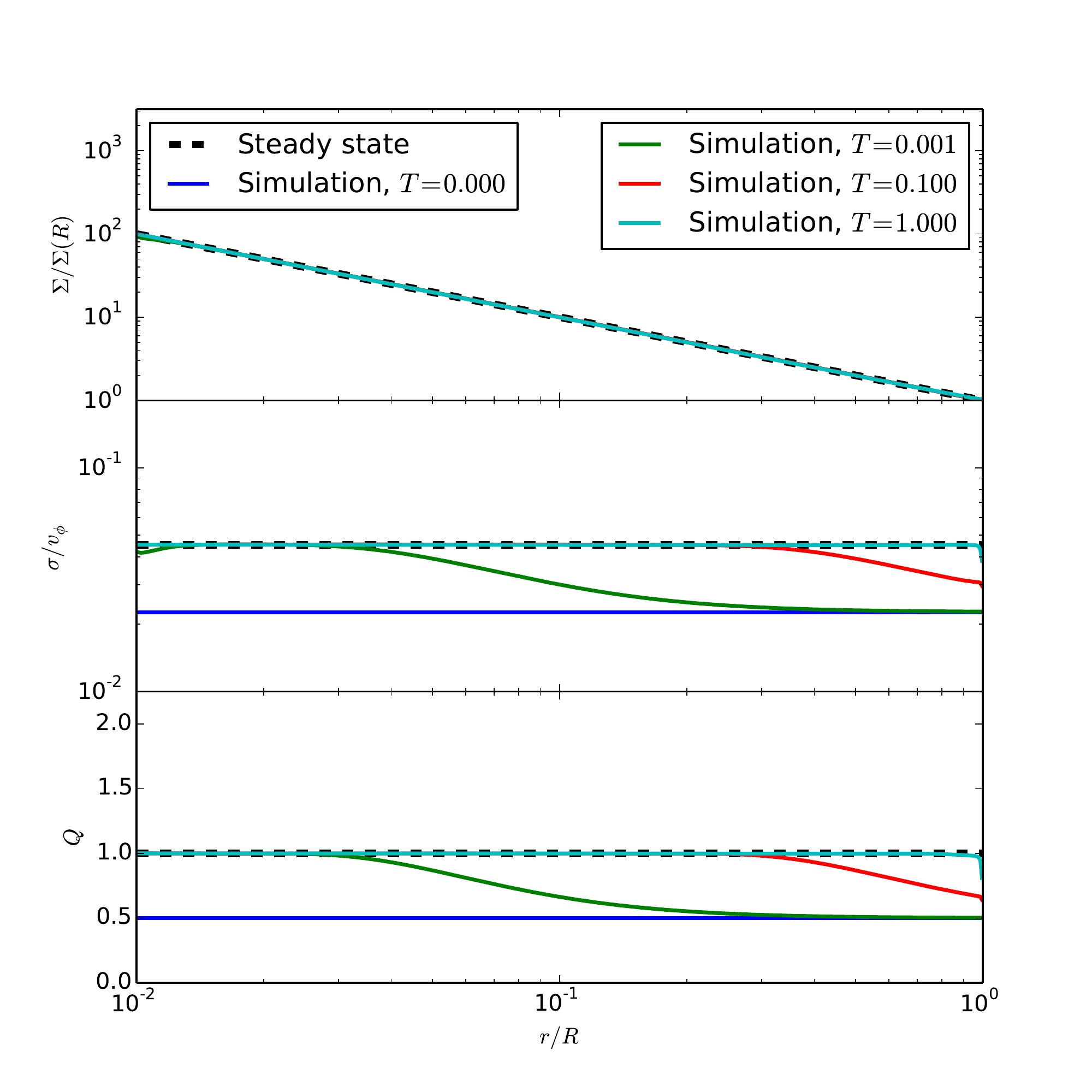}
\end{center}
\caption{
\label{fig:gidisk2}
Same as Figure \ref{fig:gidisk1}, but now showing a simulation that starts out of equilibrium with $Q=0.5$.
}
\end{figure}

Although no analytic solutions are known for systems that start away from equilibrium, a more stringent test is to start the code away from the analytic solution and verify that it approaches the steady state in a physically reasonable manner. Figure \ref{fig:gidisk2} shows the evolution for a simulation that begins the same surface density as the steady state solution (equation \ref{eq:Sigmasteady}), but a velocity dispersion a factor of 2 smaller, and thus at $Q=0.5$. The enthalpy at the outer boundary condition is also a factor of 2 below the steady-state solution value. As the Figure shows, the system rapidly evolves from the inside out to $Q=1$ due to an increase in the velocity dispersion driven by viscous transport. After $\sim 1$ outer orbital time the disk has converged to the analytic solution everywhere except in the outermost cells, where low-enthalpy material continually enters the grid and is then heated by the gravitational instability. \red{This test required $\sim 3$ seconds to complete.}

\begin{figure}
\begin{center}
\includegraphics[width=1.0\textwidth]{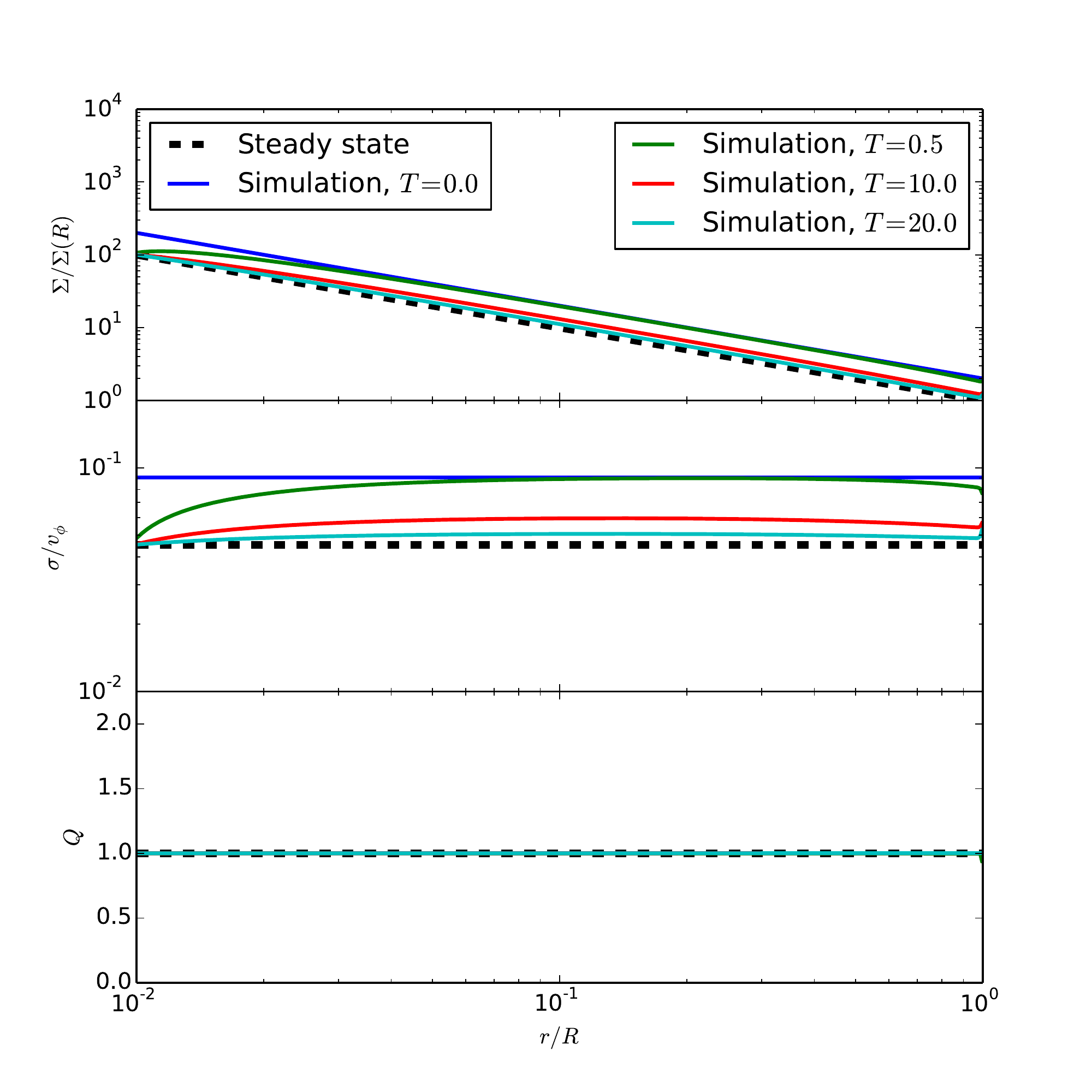}
\end{center}
\caption{
\label{fig:gidisk3}
Same as Figure \ref{fig:gidisk1}, but now showing a simulation that starts out of equilibrium with $Q=1$ but a surface density and velocity dispersion that are both double the steady-state value.
}
\end{figure}

Figure \ref{fig:gidisk3} shows the results of a test in which the system begins at $Q=1$, but with the surface density and velocity dispersion both increased by a factor of 2 relative to the steady state solution. The outer enthalpy boundary condition is equal to that of the analytic solution. Again, the system evolves toward the steady-state solution. The time required for this evolution is $\sim 10$ orbits because reaching the steady state requires decreasing the column density, and thus draining material out of the disk through the inner boundary. \red{This test required $\sim 3-4$ minutes of computation time.}

\subsection{Singular Ring with a Complex Equation of State}
\label{ssec:ringrad}

Our fourth and final test focuses on \texttt{VADER}'s ability to handle complex equations of state. As is the case for a non-equilibrium gravitational instability-dominated disk, no analytic solutions are known for this case, but we can nevertheless verify that the code gives a physically realistic solution, and that it shows good conservation properties. We therefore choose to repeat the singular ring test, but using parameters such that the ring encounters both gas pressure-dominated and radiation pressure-dominated regimes.

For simplicity consider a disk where the shapes of the vertical density and temperature distributions are fixed, so that the density and temperature can be separated in $r$ and $z$, i.e.,
\begin{eqnarray}
\rho(r,z) & = & \overline{\rho}(r) a(\zeta) \\
T(r,z) & = & \overline{T}(r) t(\zeta)
\end{eqnarray}
where $\zeta = z/z_0$ is a dimensionless height, $z_0$ is an arbitrary vertical scale factor, and the vertical distribution functions $a(\zeta)$ and $t(\zeta)$ are both normalized such that $\int a(\zeta)\, d\zeta = \int t(\zeta) \, d\zeta = 1$. The vertically-integrated column density and gas and radiation pressures are
\begin{eqnarray}
\Sigma & = &  \overline{\rho} z_0  \\
P_g & = & \frac{k_B}{\mu m_{\rm H}} \Sigma \overline{T} \int a(\zeta) t(\zeta) \, d\zeta \equiv \frac{k_B}{\mu m_{\rm H}} \Sigma T_{\rm eff}\\
P_r & = & \frac{1}{3} a \overline{T}^4 z_0 \int t(\zeta)^4\,d\zeta \equiv \frac{1}{3} f a T_{\rm eff}^4 z_0,
\end{eqnarray}
where $\mu$ is the mean molecular weight, and we have defined
\begin{eqnarray}
T_{\rm eff} & = & \overline{T} \int a(\zeta) t(\zeta) \, d\zeta \\
f & = & \frac{\int t(\zeta)^4\, d\zeta}{\left[\int a(\zeta) t(\zeta)d\zeta \right]^4}.
\end{eqnarray}
The corresponding vertically-integrated internal energies are
\begin{eqnarray}
E_{{\rm int},g} & = & \frac{P_g}{\gamma_g - 1} \\
E_{{\rm int},r} & = & 3P_r,
\end{eqnarray}
where $\gamma_g$ is the gas adiabatic index. The total pressure and internal energy are simply the sums of the gas and radiation components, i.e., $P = P_g + P_r$ and $E_{\rm int} = E_{{\rm int},g} + E_{{\rm int},r}$. From these definitions, with a bit of algebra one can show that the equation of state parameters $\gamma$ and $\delta$ can be written in terms of the total pressure and internal energy as
\begin{eqnarray}
\gamma & = & \frac{(16-3\gamma_g)P + (16-15\gamma_g)E_{\rm int}}{9P + (13-12\gamma_g) E_{\rm int}} \\
\delta & = & \frac{4(3 P - E_{\rm int})\left[(\gamma_g-1)E_{\rm int}-P\right]}{P \left[3 (\gamma_g-1)E_{\rm int}+(3\gamma_g-7)P\right]}.
\end{eqnarray}

Since the problem is not dimensionless once a real equation of state is added, we adopt the following dimensional parameters. The simulation domain has inner radius $r_{-1/2} = 1.5\times 10^{10}$ cm, outer radius $r_{N-1/2} = 1.5\times 10^{12}$ cm, and a rotation curve corresponding to Keplerian motion about a central object of mass $M=3$ $M_\odot$. The initial ring of material is located at $R_0 = 7.5\times 10^{11}$ cm, its mass and effective temperature are $M = 1.0\times 10^{-6}$ $M_\odot$ and $T_{\rm eff} = 10^4$ K, its adiabatic index $\gamma_g = 5/3$, and the scale height parameter $f z_0 = 7.5\times 10^9$ cm. The kinematic viscosity $\nu = 1.483\times 10^{11}$ cm$^2$ s$^{-1}$, which gives a characteristic evolution time $t_s = 10^4$ yr. With these choices, the ring has $P_g \gg P_r$ in its interior, but as it expands, its edges heat up to the point where $P_r \gg P_g$. The simulations use a linear grid of 4096 cells, backwards Euler updating, and piecewise constant interpolation of enthalpy; the latter two choices are made to suppress numerical ringing at the discontinuous expansion front. We also use a time step tolerance $C=1.0$ rather than $0.1$ in order to speed up the simulations. \red{The total time required to run these simulations was $\sim 5$ minutes for the case without radiation, and $\sim 3$ minutes for the case with radiation.}

\begin{figure}
\begin{center}
\includegraphics[width=1.0\textwidth]{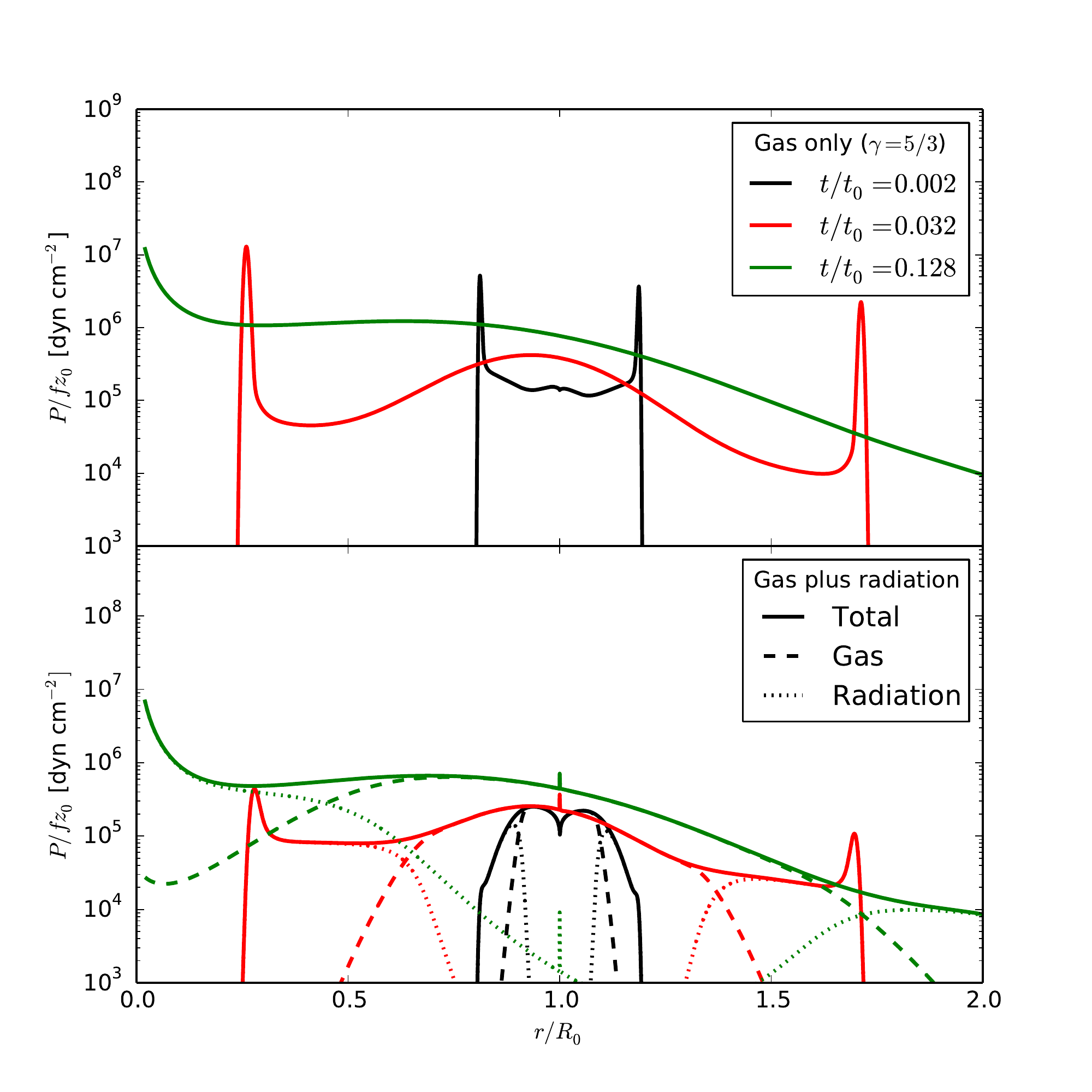}
\end{center}
\caption{
\label{fig:ringrad1}
Vertically-integrated pressure $P$ divided by the scale height parameter $fz_0$ at several times in the singular ring test. The top panel shows a simulation omitting radiation pressure ($\gamma=5/3$, $\delta=0$), while the bottom panel shows an otherwise identical simulation with a complex equation of state including radiation pressure. In the gas plus radiation run, we show both total pressure (solid lines) and the pressures due to gas (dashed lines) and radiation (dotted lines) alone.
}
\end{figure}

\begin{figure}
\begin{center}
\includegraphics[width=1.0\textwidth]{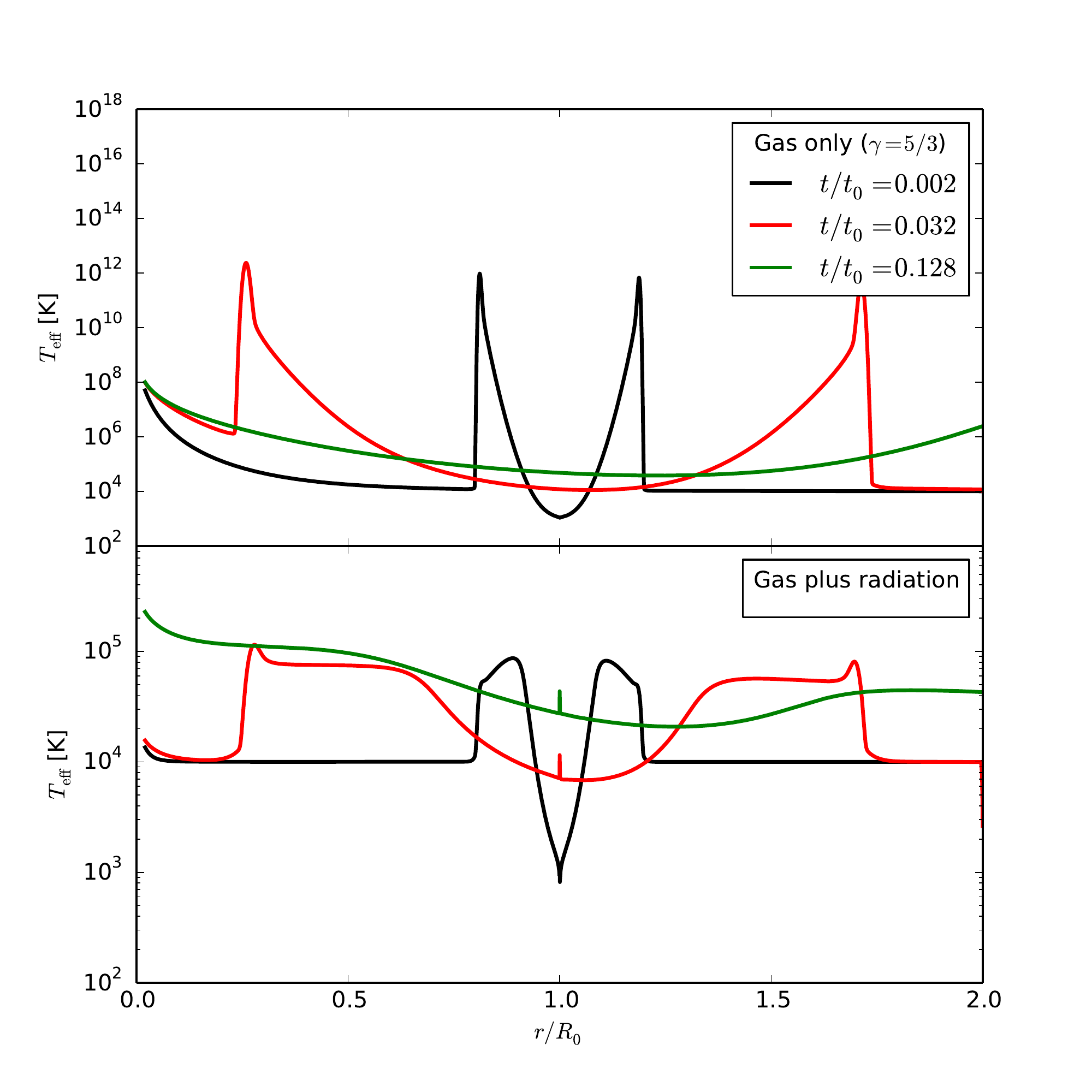}
\end{center}
\caption{
\label{fig:ringrad2}
Same as Figure \ref{fig:ringrad1}, but now showing effective temperature $T_{\rm eff}$ versus position in the singular ring test without (top) and with (bottom) radiation pressure. Notice the difference in scales between the top and bottom panels.
}
\end{figure}

Figures \ref{fig:ringrad1} and \ref{fig:ringrad2} show the pressure and temperature distributions in the simulations, both for a case where we ignore radiation pressure and use constant values $\gamma=5/3$, $\delta=0$, and a case including radiation pressure. (The column density distributions are nearly identical to those shown in Figure \ref{fig:ring1}, as is expected since the choice of viscosity for this problem is such that the column density evolution does not depend on the pressure.) As shown in the Figures, in both cases the spreading ring has sharp rises in pressure and temperature at its low-density edges, where the ring encounters near-vacuum. However, in the run including radiation pressure the pressure and temperature in these spikes are greatly reduced. In the run without radiation pressure, the temperature rises to unphysical values because the pressure must be carried entirely by the gas, and the gas pressure is linear in temperature and inversely proportional to surface density. In the run including radiation, the pressure at the ring leading and trailing edges is carried by radiation instead, and the temperature rise is vastly reduced because the radiation pressure rises as the fourth power of temperature, and does not scale with gas surface density. There is also a small spike in the pressure at the original ring location in the case with radiation pressure. It is not clear if this is physically real, or if the spike is a purely numerical effect resulting from the unresolved, singular initial condition.

This test also enables us to check the level of conservation of energy in the computation with the complex equation of state. We simulate a time interval from $t/t_s = 0$ to $0.128$, as shown in the Figures, and record the total energy in the computational domain at 65 times uniformly spaced throughout this interval. In the run with constant $\gamma$ we find that the maximum change in total energy in the computational domain, after accounting for energy transmitted across the grid boundary by advection or torques, is $9.1\times 10^{-15}$ of the initial energy, and that the mean difference between the initial energy and the energy at later times is $4.3\times 10^{-15}$ of the initial energy. This is consistent with our expectations that the algorithm should conserve total energy to machine precision. In contrast, the run including radiation pressure has maximum and mean energy conservation errors of $3.7\times 10^{-14}$ and $1.4\times 10^{-14}$ of the initial energy. Thus, while conservation is not quite at machine precision, the loss of precision is only $\sim 1$ digit of accuracy.

\section{Performance}
\label{sec:performance}

The tests presented in the previous section demonstrate that \texttt{VADER} and the algorithms on which it is based provide correct solutions to a number of problems\red{, and give a rough indication of the performance of the code}. Here we investigate the performance of the code \red{in much more detail}. We are particularly interested in the performance of the implicit solver and the Anderson acceleration code, because, while Anderson acceleration accelerates convergence and reduces the number of iterations requires in the implicit solver, it also requires a linear least squares solve that increases the cost per iteration. The tradeoff between these two is almost certainly problem-dependent, and may also be processor- and compiler-dependent, but the tests we describe here can serve as a guide for users in selecting appropriate parameters for their own problems. All the tests we discuss in this section were performed on a single core of a 2 GHz Intel i7 chip on a system running Mac OS X v.~10.9.3; \texttt{VADER} was compiled using \verb=gcc-4.8= with optimization level \verb=-O3=, while the GNU Scientific Library was built using its default options. We obtain code timing using the C \verb=clock()= function.

\begin{figure}
\begin{center}
\includegraphics[width=1.0\textwidth]{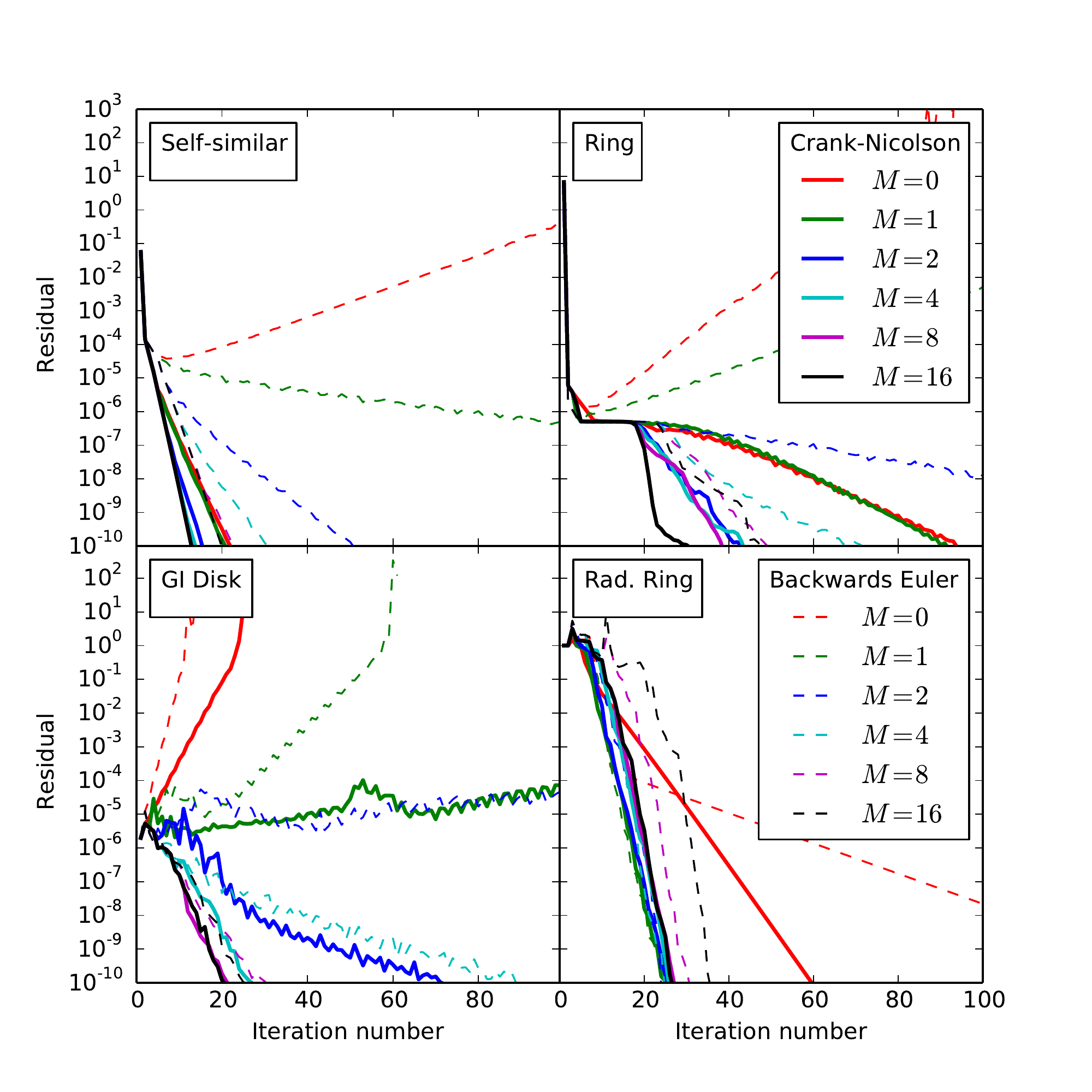}
\end{center}
\caption{
\label{fig:performance1}
Maximum normalized residual $\max|\bmath{R}_{0,i}|$ (equation \ref{eq:tol}) remaining after $N$ iterations in the four test problems described in Section \ref{sec:performance}. Solid lines show updates using the Crank-Nicolson method, and dashed lines using the backwards Euler method. Colors indicate the order $M$ of Anderson acceleration used, with $M=0$ corresponding to no acceleration (standard Picard iteration).
}
\end{figure}

We first verify that Anderson acceleration does, as expected, lead to much more rapid convergence of the iterative solver. To test this, we run each of our four test problems described in Section \ref{sec:acctests} -- the self-similar disk, the singular ring, the gravitational instability-dominated disk, and the radiation pressure ring -- for one time step, starting from the initial conditions as described in the previous Section. We use time steps of $10^{-2.5} t_s$, $10^{-6} t_s$, $10^{-3.5} t_{\rm orb}$, and $10^{-7.5} t_s$, respectively, and we test both Crank-Nicolson and backwards Euler updating. We set the tolerance on the iterative solver to $10^{-10}$, and allow a maximum of 100 iterations. Figure \ref{fig:performance1} shows how the residual changes versus number of iterations for each of these runs, and Table \ref{tab:convergence} shows the number of iterations and the wall clock time required to converge. The performance of the algorithm is in line with our expectations. We also see that the Crank-Nicolson method almost always produces faster convergence than the backwards Euler method, which is not surprising due to its higher order of accuracy.

Note that, at lower orders of Anderson acceleration, for some problems the iteration diverges rather than converging, until eventually the code produces non-numerical values (\verb=Inf= or \verb=NaN=), at which point the solver halts iteration. In a full simulation, such cases of divergence are treated exactly like cases where the solver fails to converge within the prescribed maximum number of iterations, i.e., the time step is attempted again using a reduced value of $\Delta t$ (see Section \ref{ssec:timestep}).

\begin{table}
\centering
\begin{minipage}{140mm}
\centering
\scriptsize
\begin{tabular}{ccrrrrrrrr}
\hline\hline
& & \multicolumn{2}{c}{Self-similar} & \multicolumn{2}{c}{Ring} & \multicolumn{2}{c}{GI Disk} & \multicolumn{2}{c}{Rad.~ring} \\
 & \multicolumn{1}{l}{$M$} & CN & BE & CN & BE & CN & BE & CN & BE \\ \hline \hline
\multirow{6}{*}{$N_{\rm iter}$} 
 & 0$\ldots\ldots$ & 22 & --\qquad & 94 & --\qquad & -- & --\qquad & 60 & -- \\
& 1$\ldots\ldots$ & 21 & --\qquad & 92 & --\qquad & -- & --\qquad & 26 & 25 \\
& 2$\ldots\ldots$ & 16 & 51\qquad & 44 & --\qquad & 72 & 100\qquad & 26 & 24 \\
& 4$\ldots\ldots$ & 14 & 31\qquad & 44 & 72\qquad & 27 & 89\qquad & 26 & 25 \\
& 8$\ldots\ldots$ & 13 & 23\qquad & 39 & 50\qquad & 22 & 31\qquad & 28 & 31 \\
& 16$\ldots\ldots$ & 13 & 20\qquad & 31 & 48\qquad & 21 & 26\qquad & 27 & 36 \\
\hline
\multirow{6}{*}{Time [ms]}
& 0$\ldots\ldots$ &  1.80 &  -- & 49.24 & -- &  -- &  -- & 33.48 & -- \\
& 1$\ldots\ldots$ &  2.33 & -- & 72.79 & -- & -- & -- & 21.46 & 20.40 \\
& 2$\ldots\ldots$ &  1.92 &  6.15 & 45.41 & -- & 13.34 & 18.47 & 31.14 & 25.64 \\
& 4$\ldots\ldots$ &  2.14 &  4.83 & 63.02 & 99.32 &  7.93 & 25.05 & 52.44 & 42.25 \\
& 8$\ldots\ldots$ &  2.91 &  6.12 & 128.45 & 157.75 &  7.60 & 11.51 & 123.55 & 120.10 \\
& 16$\ldots\ldots$ &  3.53 &  8.94 & 248.57 & 424.27 & 10.99 & 15.76 & 283.69 & 412.33 \\
\hline\hline
\end{tabular}
\end{minipage}
\caption{
\label{tab:convergence}
Number of iterations and total wall clock time required for convergence during one step of the self-similar disk, singular ring, gravitational instability-dominated disk, and singular ring with radiation pressure tests, using Crank-Nicolson and backwards Euler updating methods. The quantity $M$ is the Anderson acceleration parameter, with $M=0$ indicating no acceleration (standard Picard iteration). Blank entries indicate that the solver failed to converge within 100 iterations.
}
\end{table}

While this test shows that Anderson acceleration does speed convergence in terms of number of iterations, and does allow larger time steps, it does not prove that the extra computational cost per iteration is worthwhile. Indeed, carefully examining the timing results given in Table \ref{tab:convergence}, we see that the wall clock time is by no means a monotonically decreasing function of $M$, even in cases where the number of iterations is. To evaluate this question, we next run each of our test simulations for 1000 time steps or until the simulation completes, whichever comes first, and measure the total execution time. For this test we return the iterative solver tolerance to its default value of $10^{-6}$, the maximum number of iterations to its default of 40, and allow the time step to be set by the normal \texttt{VADER} procedure. We use a time step restriction $C=0.1$ for the self-similar and gravitational instability problems, and $C=1.0$ for the two ring problems.

\begin{figure}
\begin{center}
\includegraphics[width=0.9\textwidth]{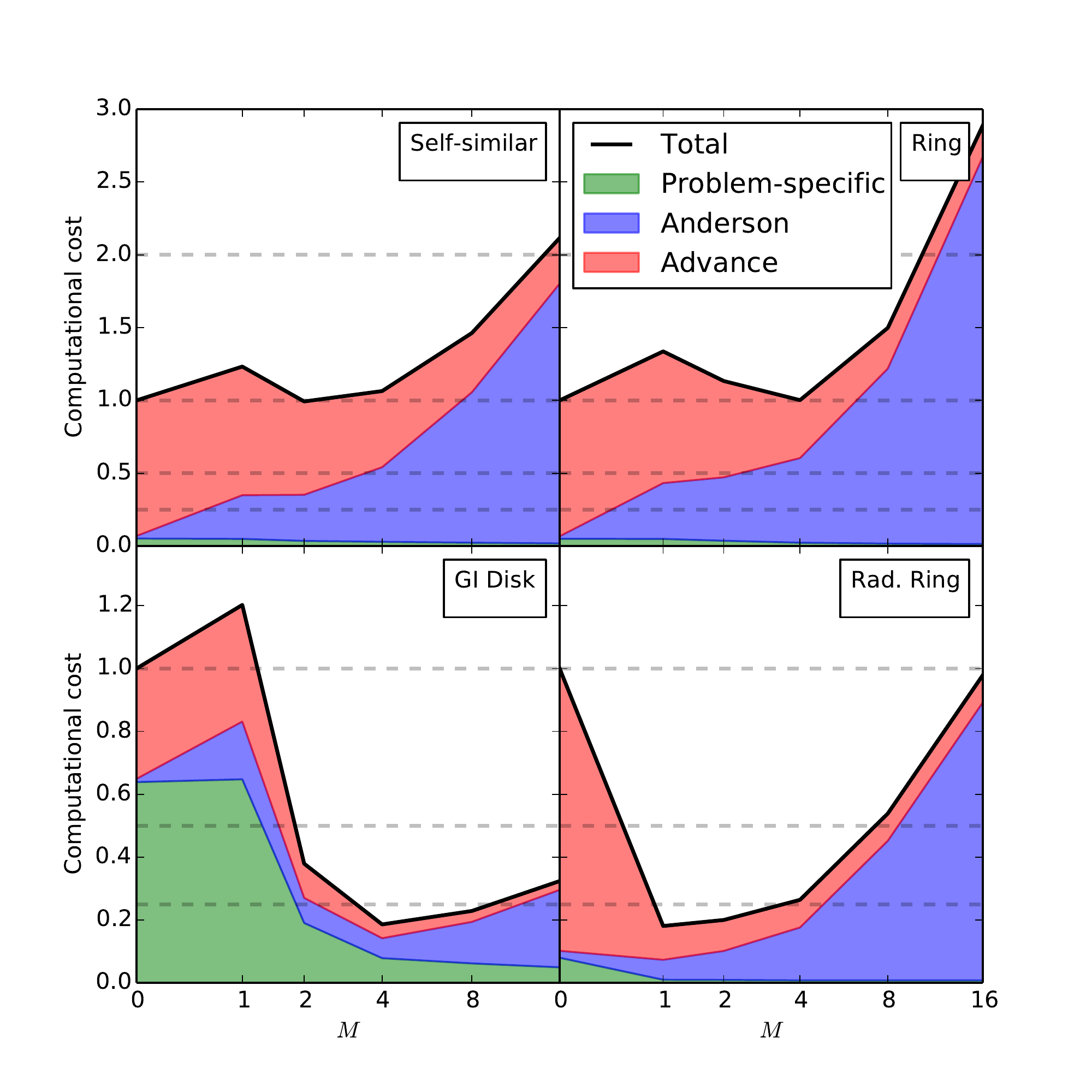}
\end{center}
\caption{
\label{fig:performance2}
Wall clock time required per unit simulation time advanced versus Anderson acceleration parameter $M$ in the four test problems, normalized so that $M=0$ corresponds to unity. Thus values $<1$ indicate a reduction in computational cost relative to the unaccelerated case, while values $>1$ indicate a slowdown. The thick black lines indicate the total cost, and shaded regions indicate the fraction of the cost contributed by different parts of the computation. Problem-specific routines, including those used to compute the viscosity ($\alpha$), source terms ($\dot{\Sigma}_{\rm src}$, $\dot{E}_{\rm int,src}$), equation of state terms ($\gamma$, $\delta$), and boundary conditions are shown in green. The Anderson acceleration step is shown in blue; for $M=0$ this is simply the cost of copying temporary arrays. The remainder of the time step advance procedure is shown in red. The dashed horizontal lines indicate values of $0.25$, $0.5$, $1.0$, and $2.0$. Note the change in $y$-axis range between the top two and bottom two panels. The plots shown are for Crank-Nicolson; the results for backwards Euler are qualitatively similar.
}
\end{figure}

In Figure \ref{fig:performance2} we plot the wall clock time required per unit of simulation time advanced in each of our four test cases. As the plot shows, the optimal Anderson acceleration parameter, and the range of cases for which it is helpful, depends strongly on the problem. For the self-similar disk and ring problems, Anderson acceleration is neutral at small $M$ and actually harmful at larger $M$. This is because the reduction in number of iterations is more than offset by the increased cost per iteration. On the other hand, for the gravitational instability-dominated disk and ring with radiation pressure problems, Anderson acceleration with $M$ of a few provides very significant gains in performance, reducing the computational cost by a factor of $\sim 5$. 

The difference in performance between the cases where Anderson acceleration helps and those where it does not arises mostly from the complexity of the implicit update. In the self-similar disk and ring problems, the viscosity and boundary conditions are trivial to compute and there are no source terms. As a result, the linear least squares solve required by Anderson acceleration contributes significantly to the total implicit update cost, and begins to dominate it for higher values of $M$. In contrast, the gravitational instability-dominated disk and radiation pressure ring problems have much higher computational costs per update. For the former, the cost is high because there is a source term and because computing the viscosity requires solving a tridiagonal matrix equation at every iteration; this is indicated by the green region in Figure \ref{fig:performance2}. For the latter there are no source terms, but because the problem uses a complex equation of state, there is additional computational work associated with updating the internal energy equation (which is included in the red region in Figure \ref{fig:performance2}).

The implications of this analysis are that the choice of optimal Anderson acceleration parameter is likely to depend on the nature of the model being used to generate the viscosity, source terms, boundary conditions, and equation of state. If these are given by simple analytic formulae, then Anderson acceleration is probably of very limited use. The more computationally complex they are to evaluate, however, the greater the advantage that one gains by using Anderson acceleration to reduce the number of iterations required.

\section{Summary \red{and Future Prospects}}
\label{sec:summary}

Thin, axisymmetric accretion disks where the transport of mass and angular momentum is approximated as being due to viscosity represent an important class of models in theoretical astrophysics. They are widely used in situations where full two- or three-dimensional simulations would be prohibitively expensive, either due to the number of orbits that would have to be simulated, or because of the dynamic range in spatial resolution that would be required. We have developed a new, extremely flexible and general method for simulating the evolution of such models. Our discretization of the equations is conservative \red{to machine precision}, and allows complete freedom in the specification of rotation curves, equations of state, forms of the viscosity, boundary conditions, and sources and sinks of energy and mass. The core of our method is an unconditionally-stable update strategy that uses accelerated fixed point iteration to achieve rapid convergence. We show that this technique allows relatively large time steps, and that it significantly reduces the number of implicit iterations required to advance the simulation a specified time. \red{In practical tests, even at resolutions as high as $4096$ cells, and using complex equations of state or disk $\alpha$ parameters that must themselves be computed iteratively, the code is can evolve a disk for many viscous evolution timescales in no more than a few minutes of computation time on a single CPU. In tests with simple equations of state and fixed $\alpha$, computational times are only a few seconds.} We have implemented our algorithm in a new open source code called the Viscous Accretion Disk Evolution Resource \texttt{VADER}, which can be downloaded from \url{https://bitbucket.org/krumholz/vader/}. The code is designed for modularity, so that users can easily implement their own models for viscosity and similar parameters that control disk evolution.

\red{While the number of potential applications for \texttt{VADER} is large, we end this discussion by highlighting a few possible examples. One, already underway (Krumholz \& Kruijssen, 2015, in preparation), is to model the long-term behavior of the gas around the central black in the Milky Way and other galaxies. Observations suggest that the gas accumulates over long timescales before undergoing periodic starburst events \citep[e.g.][]{kruijssen15a}, and this process can be modeled as gas undergoing slow viscous accretion before accumulating to the point where it becomes gravitationally unstable and undergoes a starburst. The customization required for this project consists mostly of implementing a custom rotation curve, viscosity representing the instabilities that likely drive the gas migration, and testing for the onset of gravitational instability.}

\red{In the context of planet formation, viscous evolution models have been used to study the long-term interaction of planets with the disks out of which they form, and the migration of the planets through viscous disks \citep[e.g.,][]{lyra10a, horn12a}. \texttt{VADER} is well-suited for this application, and could be customized to it simply by implementing cooling and viscosity prescriptions appropriate to a protoplanetary disk, by coupling the state of the disk found by \texttt{VADER} to a calculation of the torques on planets, and perhaps by modifying the viscosity prescription to incorporate the back-reaction of disk-planet torques on the transport of gas through the disk. Also in this context, \texttt{VADER} could be used to simulate the photoevaporation of disks around young stars by high-energy radiation, either from the central star or from an external source \citep[e.g.][]{adams04a, gorti09a, gorti09b}. In this case one could treat the effects of photoevaporation by adding mass and energy source terms to represent the rates of mass loss and heating driven by ultraviolet and X-ray photons striking the disk.}

\red{For accretion disks around compact objects, a number of authors have used viscous disk models to study variability and flaring on timescales associated with viscous evolution \citep[e.g.,][]{cambier13a}. \texttt{VADER} is well-suited to this problem too, and could be customized to it by adding in prescriptions for viscosity and radiative heating and cooling, and by post-processing the \texttt{VADER} models to predict observable X-ray fluxes. One could also add mass and energy source terms representing the exchange of mass and energy with a hot corona.}

\red{This list of potential applications is certainly not exhaustive, but its breadth and diversity should make clear that a general-purpose viscous disk evolution code is a tool of wide applicability.}  The availability of such a code should reduce the need for every modeler to develop his or her own approach to a standard problem.

\section*{Acknowledgements}

We thank D.~Kruijssen for stimulating discussions that helped inspire this work, and the authors and maintainers of the \href{http://www.gnu.org/software/gsl/}{GNU Scientific Library} for providing a valuable resource. \red{We thank W.~Lyra and T.~Tanaka for helpful comments on the version of this paper that was posted as arXiv:1406.6691v1, the two anonymous referees for useful reports.} MRK and JCF acknowledge support from NSF CAREER grant AST-0955300. MRK thanks the Aspen Center for Physics, which is supported by NSF Grant PHY-1066293, for providing the forum during which this work was begun.

\bibliographystyle{model2-names}
\bibliography{refs}

\begin{appendix}

\section{Tabulated Rotation Curves}
\label{app:tabrotation}

\texttt{VADER} allows arbitrary rotation curves $v_\phi(r)$, and these can be specified either via user-supplied analytic functions, or in the form of a table of $(r, v_\phi)$ values. In the latter case \texttt{VADER} generates a rotation curve $v_\phi$ on the grid via interpolation, and the potential $\psi$ by integrating the interpolating function. However, the interpolation procedure requires special care to ensure smoothness. Equations (\ref{eq:masscons}) and (\ref{eq:encons}) involve the second derivative of the torque $\mathcal{T}$, which is proportional to the rotation curve index $\beta = \partial\ln v_\phi/\partial\ln r$. Thus derivatives up to $\partial^3 \ln v_\phi/\partial \ln r^3$ appear in the evolution equations, and it is therefore highly desirable, for both computational stability and to avoid imposing artifacts on the solution, that the interpolating function constructed to approximate a table of $(r, v_\phi)$ values have at least three continuous derivatives.

To achieve this aim, \texttt{VADER} constructs tabulated rotation curves using basis splines (b-splines). In the b-spline method, the domain of the function to be fit is broken up into a set of intervals, separated by breakpoints, and one must choose the degree $D$ of the fit, the number of breakpoints $B$, and their locations. For the choice of locations, we use the method of \citet{gans84a}, who show that errors in the resulting fit are minimized if the breakpoints are distributed evenly in the size of the the interval weighted by the square root of the function being fit. Let $(r_n, v_{\phi,n})$ be our table of $N$ input data, ordered from $n=0 \ldots N-1$ with $r_n$ increasing monotonically, and let $b_m$ be the location of the $m$th breakpoint, $m = 0\ldots B-1$. We set $b_0 = r_0$ and $b_{B-1} = r_{N-1}$, and we assign the remaining breakpoints $b_m$ via the following algorithm. Let
\begin{equation}
S_T \equiv \frac{1}{B+1} \sum_{n=0}^{N-1} v_{\phi,n}^{1/2} dx_n,
\end{equation}
where $dx_n = (1/2) \log (r_{n+1}/r_{n-1})$ for $n \neq 0, N-1$, $dx_0 = \log(r_1/r_0)$, and $dx_{N-1} = \log(r_{N-1}/r_{N-2})$. Starting from a breakpoint $b_m$ located at data value $r_{n_m}$, we set the position of the next breakpoint to $b_{m+1} = r_{n_{m+1}}$, where $n_{m+1}$ is the smallest index for which
\begin{equation}
\sum_{n=n_m+1}^{n_{m+1}} v_{\phi,n}^{1/2} dx_n \geq S_T.
\end{equation}
This assures that the breakpoints are as uniformly distributed as possible following the criterion of \citet{gans84a}.\footnote{If the number of points $N$ is very small, it is conceivable that, due to the discrete placement of the data points $r_n$, this algorithm might result in not all breakpoints $b_m$ being assigned before we reach the end of the array. In this case we end up with the last breakpoint interval being of size 0, i.e., $b_{B-2} = b_{B-1}$. However, this presents no obstacles to the remainder of the algorithm.} Once the breakpoint locations are chosen, the function and its derivatives can be constructed by the standard basis spline method.

\begin{figure}
\begin{center}
\includegraphics[width=0.5\textwidth]{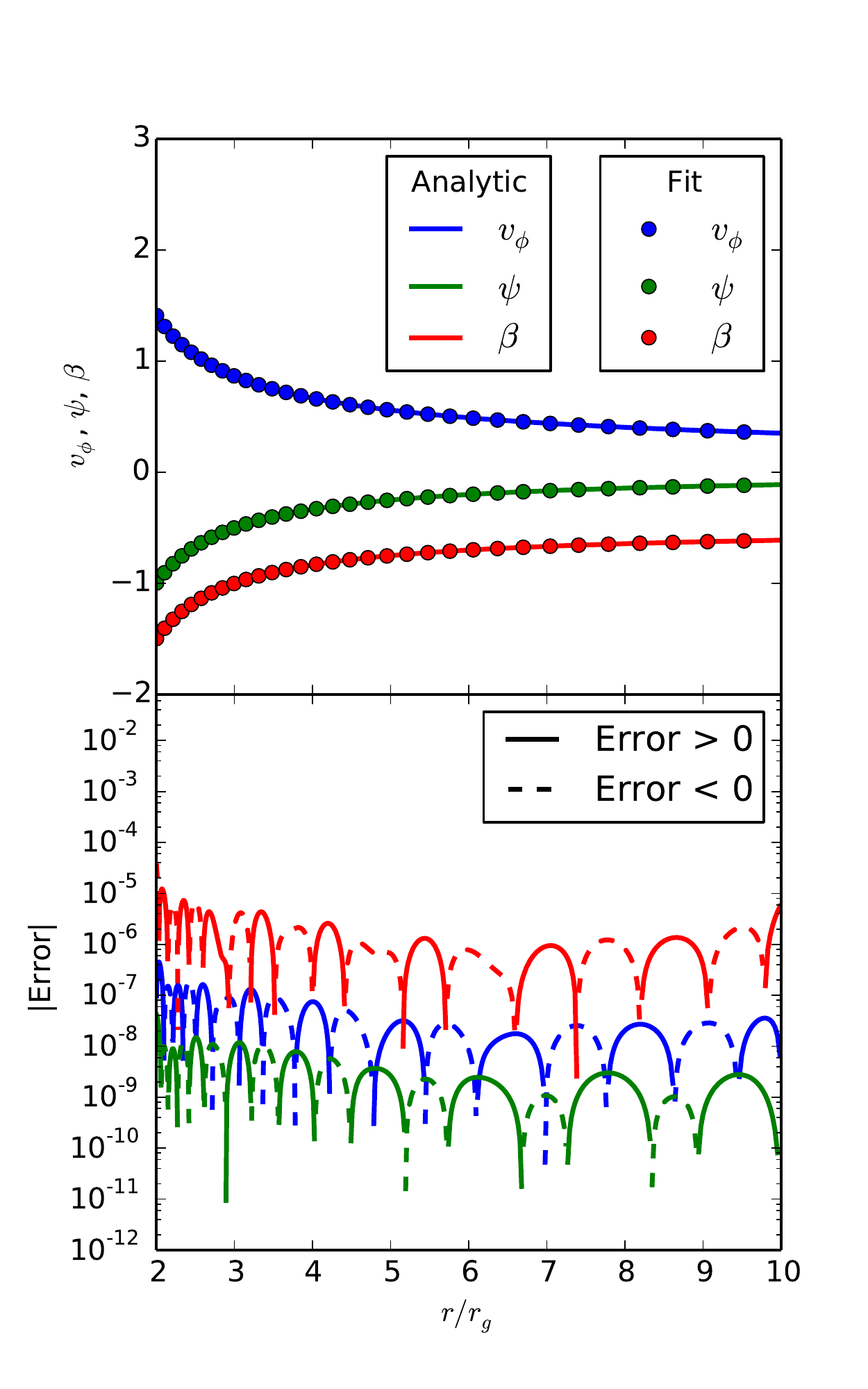}
\end{center}
\caption{
\label{fig:rotcurve1}
B-spline reconstruction of the rotation curve for a \citet{paczynsky80a} potential using a fit of degree $D=6$ with $B=15$ breakpoints. In the upper panel, we show the exact analytic values of $v_\phi$, $\psi$, and $\beta$ (solid lines) and the numerical fits (data points, only every 16th point shown for clarity). All quantities are plotted in units where $GM = r_g = 1$, and the gauge of the potential set so that $\psi \rightarrow 0$ as $r\rightarrow \infty$. In the bottom panel, we plot the relative error of the b-spline reconstruction versus $r$, defined as $\mathrm{Error} = (v_{\phi,{\rm fit}}-v_{\phi,{\rm exact}})/v_{\phi,{\rm exact}}$, and similarly for $\psi$ and $\beta$.
}
\end{figure}

To demonstrate and evaluate the performance of this capability, we perform two tests. For the first, we take the \citet{paczynsky80a} approximation for a black hole potential
\begin{equation}
\psi_{\rm PW} = \frac{GM}{r - r_g},
\end{equation}
where $r_g = 2 G M/c^2$ is the horizon radius, and generate a table of rotation velocities by analytically evaluating $v_\phi = d\psi_{\rm PW}/dr$ at points from $1.9r_g - 10.1r_g$, spaced in units of $0.1rg_g$. We then use this table to generate a 6th-order b-spline fit using 15 breakpoints, and from that fit compute the potential $\psi$ and the rotation curve index $\beta$ on a grid of 512 points logarithmically spaced from $r = 2-10 r_g$. Figure \ref{fig:rotcurve1} shows the results of the fit as compared with the analytic values for $\psi$, $v_\phi$, and $\beta$. Clearly the b-spline reconstruction is excellent.

\begin{figure}
\begin{center}
\includegraphics[width=0.5\textwidth]{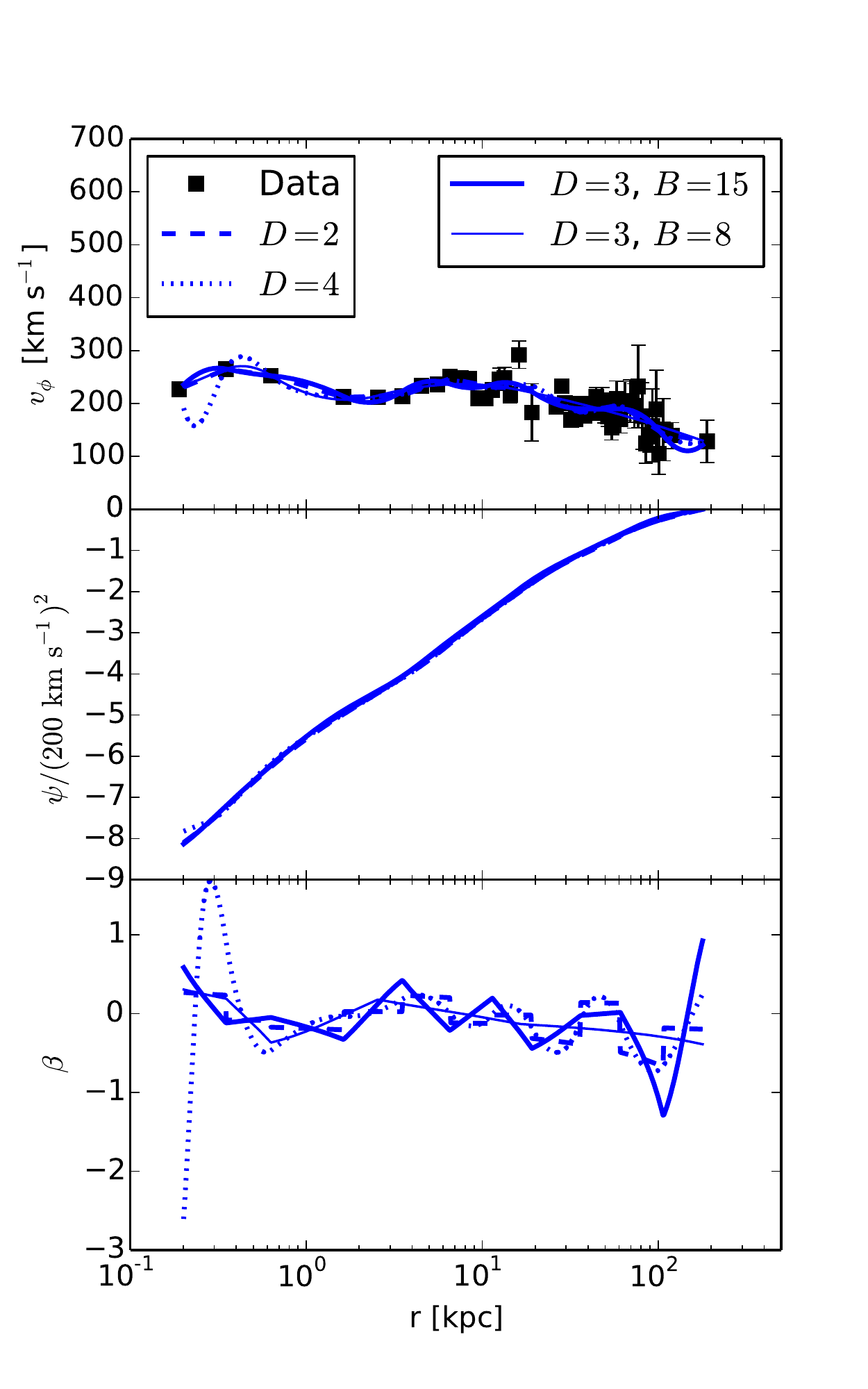}
\end{center}
\caption{
\label{fig:rotcurve2}
B-spline reconstruction of the rotation curve of the Milky Way. The top panel shows data from \citet{bhattacharjee14a} (black points with error bars), together with b-spline fits of degree $D=2$, 3, and 4 (blue lines). The $D=2$ and $D=4$ fits use 15 breakpoints each, while for $D=3$ we show models with both $B=8$ and $B=15$ breakpoints. The middle panel shows the potential, with a gauge chosen so that $\psi=0$ at the edge of the grid. The bottom panel shows $\beta$.
}
\end{figure}

For the second test we use a much noisier data set: a compilation of data on the rotation curve of the Milky Way from \citet{bhattacharjee14a}, using their model where the Sun is 8.5 kpc from the Galactic center and the rotation velocity at the Solar circle is 220 km s$^{-1}$. Figure \ref{fig:rotcurve2} shows the data and a variety of \texttt{VADER} b-spline reconstructions of $v_\phi$, $\psi$, and $\beta$ on a logarithmic grid of 512 points uniformly spaced from $0.2-180$ kpc. In this case it is clear that the rotation curve and potential are very well reconstructed and that fits to them do not depend strongly on the choice of degree $D$ and number of breakpoints $B$, except that $D\geq 4$ introduces artificial ringing at small radii. In contrast, $\beta$ does depend at least somewhat on these choices. For this particular data set, there appears to be no value of $D$ that both guarantees that the derivative of $\beta$ be continuous and that $\beta$ itself remains within physically-reasonable values (roughly $-0.5$ to $1$). With data sets of this sort, one must either accept discontinuities in the derivative of $\beta$, prune the data set more carefully, or use a more sophisticated fitting procedure.

\end{appendix}

\end{document}